\documentclass[journal]{IEEEtran}
\usepackage{setspace}
\usepackage{amsmath}
\usepackage{amsfonts}
\usepackage{cjk}
\usepackage{indentfirst}
\usepackage{cite}
\usepackage{bm}
\usepackage{graphicx}
\usepackage{subfigure}
\usepackage{booktabs}
\usepackage{multirow}
\begin{document}

\title{Nonconvex Regularization Based Sparse Recovery and Demixing with Application to Color Image Inpainting}

\author{Fei Wen, \textit{Member}, \textit{IEEE},
        Lasith Adhikari, \textit{Student Member}, \textit{IEEE},
        Ling Pei, \textit{Member}, \textit{IEEE},
        Roummel F. Marcia, \textit{Member}, \textit{IEEE},
        Peilin Liu, \textit{Member}, \textit{IEEE}, and
        Robert C. Qiu, \textit{Fellow}, \textit{IEEE}

%\thanks{This work was supported in part by the National Natural Science Foundation of China (NSFC) under grants 61401501 and 61472442.}
\thanks{F. Wen, L. Pei, P. Liu and R. C. Qiu are with the Department of Electronic Engineering, Shanghai Jiao Tong University, Shanghai 200240, China (e-mail: wenfei@sjtu.edu.cn; ling.pei@sjtu.edu.cn; liupeilin@sjtu.edu.cn; rcqiu@sjtu.edu.cn).}% <-this % stops a space
%\thanks{F. Wen is also with the Air Control and Navigation Institution, Air Force Engineering University, Xian 710000, China.}
\thanks{L. Adhikari and R. F. Marcia are with Applied Mathematics, University of California, Merced, CA 95343, USA (e-mail: ladhikari@ucmerced.edu; rmarcia@ucmerced.edu).}
% <-this % stops a space
}

\markboth{}
{Shell \MakeLowercase{\textit{et al.}}: Bare Demo of IEEEtran.cls for Journals}

\maketitle

\begin{abstract}
This work addresses the recovery and demixing problem of signals that are sparse in some general dictionary. Involved applications include source separation, image inpainting, super-resolution, and restoration of signals corrupted by clipping, saturation, impulsive noise, or narrowband interference. We employ the $\ell_q$-norm ($0 \le q < 1$) for sparsity inducing and propose a constrained $\ell_q$-minimization formulation for the recovery and demixing problem. This nonconvex formulation is approximately solved by two efficient first-order algorithms based on proximal coordinate descent and alternative direction method of multipliers (ADMM), respectively. The new algorithms are convergent in the nonconvex case under some mild conditions and scale well for high-dimensional problems. A convergence condition of the new ADMM algorithm has been derived. Furthermore, extension of the two algorithms for multi-channels joint recovery has been presented, which can further exploit the joint sparsity pattern among multi-channel signals. Various numerical experiments showed that the new algorithms can achieve considerable performance gain over the $\ell_1$-regularized algorithms.
\end{abstract}


\IEEEpeerreviewmaketitle {}

\section{Introduction}
\label{sec:intro}

This work considers the problem of identifying two sparse vectors ${{\bf{x}}_k} \in {\mathbb{R}^{{n_k}}}$, $k=1,2$, from the linear measurements ${\bf{y}} \in {\mathbb{R}^m}$ modeled as
\begin{equation}   %%(1)
{\bf{y}} = {{\bf{A}}_{\rm{1}}}{{\bf{x}}_{\rm{1}}}{\rm{ + }}{{\bf{A}}_{\rm{2}}}{{\bf{x}}_{\rm{2}}}
\end{equation}
where ${{\bf{A}}_k} \in {\mathbb{R}^{m \times {n_k}}}$ are known deterministic dictionaries. The objective is to recover and demix the two sparse signals ${{\bf{x}}_{\rm{1}}}$ and ${\bf{x}}_{\rm{2}}$ by exploiting their sparsity structure. Important application examples involving such a recovery and demix problem arise in the following scenarios.

\textit{1) Source separation}: In many applications such as the separation of texture in images [1], [2] and the separation of neuronal calcium transients in calcium imaging [3], the task is to demix the two distinct components entangled within $\bf{y}$. In this case, ${\bf{A}}_1$ and ${\bf{A}}_1$ are two dictionaries allowing for sparse representation of the two distinct features, and ${\bf{x}}_{\rm{1}}$ and ${\bf{x}}_{\rm{2}}$ are the corresponding (sparse or approximately sparse) coefficients describing these features [4]--[6].
\textit{2) Super-resolution and inpainting}: In the super-resolution and inpainting problem for images, audio, and video signals [7]--[9], only a subset of the entries of the desired signal ${{\bf{y}}_0} = {{\bf{A}}_{\rm{1}}}{{\bf{x}}_{\rm{1}}}$ is available. The task is to fill in the missing parts in ${{\bf{y}}_0}$ from ${\bf{y}}$ . In this case, ${{\bf{A}}_2} = {{\bf{I}}_m}$ and ${\bf{x}}_{\rm{2}}$ accounts for the missing parts of the desired signal.
\textit{3) Interference cancellation}: In many audio, video, or communication applications, it is desired to recover a signal corrupted by narrowband interference, such as electric hum [5]. Such interference can be naturally sparsely represented in the frequency domain. In this case, ${\bf{A}}_2$ is an inverse discrete Fourier transform matrix allowing for sparse representation of the interference.
\textit{4) Saturation and clipping restoration}: In practical systems where the measurements are quantized to a finite number of bits, nonlinearities in amplifiers may result in signal saturation, which causes significant nonlinearity and potentially unbounded errors [5], [10], [11]. In this situation, the task is to restore ${{\bf{y}}_0} = {{\bf{A}}_{\rm{1}}}{{\bf{x}}_{\rm{1}}}$ from its situated measurement $\bf{y}$, with ${\bf{x}}_2$ represents the saturation errors.
\textit{5) Robust recovery in impulsive noise}: In practical applications, impulsive noise may come from missing data in the measurement process, transmission problems [12]--[14], faulty memory locations [15], buffer overflow [16], reading out from unreliable memory, and has been raised in many image and video processing works [17]--[19]. In this case, ${{\bf{A}}_2} = {{\bf{I}}_m}$ and ${{\bf{x}}_{\rm{2}}}$ represents the (sparsely) impulsive noise, and the task is to recover the sparse signal ${{\bf{x}}_{\rm{1}}}$ from $\bf{y}$.

In all these applications, ${{\bf{x}}_{\rm{1}}}$ and ${{\bf{x}}_{\rm{2}}}$ in model (1) can be reasonably assumed to be sparse. To recover ${{\bf{x}}_{\rm{1}}}$ and ${{\bf{x}}_{\rm{2}}}$ from $\bf{y}$, we use $\ell_q$-norm with $0\le q<1$ for sparsity promotion and propose the following formulation
\begin{equation}   %%(2)
\mathop {\min }\limits_{{{\bf{x}}_1},{{\bf{x}}_2}} \left\{ {\mu \left\| {{{\bf{x}}_1}} \right\|_{{q_1}}^{{q_1}} + \left\| {{{\bf{x}}_2}} \right\|_{{q_2}}^{{q_2}}} \right\}
~~\mathrm{subject~to}~~
{{\bf{A}}_1}{{\bf{x}}_1} + {{\bf{A}}_2}{{\bf{x}}_2} = {\bf{y}}
\end{equation}
where $0 \!\le\! {q_{\rm{1}}},{q_2} \!<\! 1$, $\mu$ is a positive parameter which takes the statistic difference between the two components into consideration and its optimal value is related with the statistical information of the true signals ${{\bf{x}}_{\rm{1}}}$ and ${{\bf{x}}_{\rm{2}}}$, ${\| \cdot \|_q}$ is the $\ell_q$ quasi-norm defined as ${\| {\bf{v}} \|_q} = {(\sum\nolimits_{i = 1}^n {{{| {{v_i}} |}^q}} )^{1/q}}$.

To achieve sparsity inducing, the $\ell_1$-norm regularization is the most widely used technique since an $\ell_1$-minimization problem is tractable due to its convexity.
However, the $\ell_1$-regularization has a bias problem as it would produce biased estimates for large coefficients. Meanwhile, it cannot recover a signal with the least measurements [20]. These problems can be ameliorated by using a nonconvex regularization function, such as $\ell_q$-norm or smoothly clipped absolute deviation (SCAD) [21].

Compared with $\ell_1$-regularization, $\ell_q$-regularization with $q<1$ can yield significantly better recovery performance in many applications [22]--[33], [51].
Extensive studies in compressive sensing (CS) have demonstrated that, relative to $\ell_1$-regularized sparse recovery methods, $\ell_q$-regularized methods require fewer measurements to achieve reliable reconstruction while require weaker sufficient conditions for reliable reconstruction.
More specifically, it has been shown in [20] that under certain restricted isometry property (RIP) conditions of the sensing matrix, $\ell_q$-regularized algorithms require fewer measurements to gain a good recovery than $\ell_1$-regularized ones. Moreover, the sufficient conditions in terms of RIP for $\ell_q$-minimization are weaker than those for $\ell_1$-minimization [22], [33].

\subsection{Connections to Related Work}
When ${q_1} = {q_2} = 1$, the formulation (2) becomes
\begin{equation}   %%(3)
\mathop {\min }\limits_{{{\bf{x}}_1},{{\bf{x}}_2}} \left\{ {\mu {{\left\| {{{\bf{x}}_1}} \right\|}_1} + {{\left\| {{{\bf{x}}_2}} \right\|}_1}} \right\}
~~\mathrm{subject~to}~~
{{\bf{A}}_1}{{\bf{x}}_1} + {{\bf{A}}_2}{{\bf{x}}_2} = {\bf{y}}
\end{equation}
which has been considered in [6] for source separation. When $\mu  = 1$ and ${q_1} = {q_2} = 1$, the formulation (2) degenerates to the basis-pursuit form considered in [4] for the applications of source separation, super-resolution and inpainting, interference cancellation, and robust sparse recovery.

When ${{\bf{A}}_2} = {{\bf{I}}_m}$ and ${q_1} = {q_2} = 1$, the formulation (2) can be expressed as
\begin{equation}   %%(4)
\mathop {\min }\limits_{{{\bf{x}}_1},{{\bf{x}}_2}} \left\{ {\mu {{\left\| {{{\bf{x}}_1}} \right\|}_1} + {{\left\| {{{\bf{x}}_2}} \right\|}_1}} \right\}
~~\mathrm{subject~to}~~
{{\bf{A}}_1}{{\bf{x}}_1} + {{\bf{x}}_2} = {\bf{y}}.
\end{equation}
In this case, it in fact reduces to the well-known $\ell_1$-regularized least-absolute ($\ell_1$-LA) problem for robust sparse recovery [34]
\begin{equation}   %%(5)
\mathop {\min }\limits_{{{\bf{x}}_1}} \left\{ {\mu {{\left\| {{{\bf{x}}_1}} \right\|}_1} + {{\left\| {{{\bf{A}}_1}{{\bf{x}}_1} - {\bf{y}}} \right\|}_1}} \right\}.
\end{equation}
In compressive sensing, this formulation has showed considerable gain over the $\ell_2$-loss based ones in the presence of impulsive measurement noise.
Meanwhile, for ${{\bf{A}}_2} = {{\bf{I}}_m}$, ${q_1} = 1$ and $0 \le {q_2} < 2$, the formulation (2) reduces to the robust sparse recovery formulation considered in [35], [45].
Moreover, the $\ell_q$-regularized least-squares sparse recovery methods [28]--[30] can be viewed as special cases of (2) with ${{\bf{A}}_2} = {{\bf{I}}_m}$, $0 \le {q_1} < 1$ and ${q_2} = 2$.

For the formulation (2) with $0 \le {q_{\rm{1}}},{q_2} < 1$, since both terms in the objective are nonsmooth and nonconvex, it is more difficult to solve compared with those in the above works.

\subsection{Contributions}
Generally, the constrained ${\ell _{{q_1}}}-{\ell _{{q_2}}}$ mixed minimization problem (2) is difficult to solve. The efficient alternative direction method of multipliers (ADMM) framework can be directly used to solve (2) [36], but this directly extended ADMM algorithm often fails to converge in empirical experiments (see Fig. 1 in section V). The main contributions of this work are as follows.

First, to derive convergent algorithms for (2), we propose two first-order algorithms to solve an approximation of (2) based on the block coordinate descent (BCD) and ADMM frameworks, respectively. Both algorithms are convergent under some mild conditions and scale well for high-dimensional problems. Furthermore, a sufficient condition of convergence for the proposed ADMM algorithm has been derived.

Second, to exploit the feature correlation among multi-channels of color images, the new algorithms have been extended for multi-channel joint recovery, which can achieve further performance gain in color image recovery.

Finally, we have evaluated the new algorithm via various experiments. The results showed that, with properly selected $q_1<1$ and $q_2<1$, the new algorithms can achieve considerable performance improvement over the $\ell_1$-minimization algorithms.

Matlab codes for reproducing the results in this work are available at https://github.com/FWen/Lq-Sparse-Recovery.git.

\subsection{Outline and Notations}
Section II introduces the proximity operator for the $\ell_q$-norm function, which is employed in the proposed algorithms. In Section III, the two new algorithms are presented. Section IV extends the new algorithms to the multitask case. Section V provides experimental results on image inpainting. Finally, Section VI ends the paper with concluding remarks.

\textit{Notations}: For a matrix ${\bf{M}}$, ${\| {\bf{M}} \|_F}$ is the Frobenius norm, ${\lambda _{\max }}({\bf{M}})$ and ${\lambda _{\min }}({\bf{M}})$ denote the maximal and the minimal eigenvalues of ${\bf{M}}$, respectively. $\langle {\cdot, \cdot} \rangle$ and ${(\cdot)^T}$ stand for the inner product and transpose, respectively. $\nabla f(\cdot)$ and $\partial f(\cdot)$ stand for the gradient and subdifferential of the function $f$, respectively. ${\rm{sign}}(\cdot)$ denotes the sign of a quantity with ${\rm{sign}}(0){{ = }}0$. $\bf{I}$ stands for an identity matrix with proper size. $\|\cdot\|_q$ with $q\ge0$ denotes the $\ell_q$-norm defined as ${\| {\bf{x}} \|_q} = {(\sum\nolimits_{i = 1}^{} {{{| {{x_i}} |}^q}} )^{1/q}}$. ${\rm{dist}}({\bf{x}},S): = \inf \{ {\| {{\bf{y}} - {\bf{x}}} \|_2}:{\bf{y}} \in S\} $ denotes the distance from a point ${\bf{x}} \in {\mathbb{R}^n}$ to a subset $S \subset {\mathbb{R}^n}$.

\section{Proximity Operator for $\ell_q$-Norm Function}

This section introduces the proximity operator of the $\ell_q$-norm function, which is defined as
\begin{equation}   %%(6)
{\rm{pro}}{{\rm{x}}_{q,\eta }}({\bf{t}}) = \arg \mathop {\min }\limits_{\bf{x}} \left\{ {\left\| {\bf{x}} \right\|_q^q + \frac{\eta }{2}\left\| {{\bf{x}} - {\bf{t}}} \right\|_2^2} \right\}
\end{equation}   %%
for ${\bf{x}} \in {\mathbb{R}^m}$, and with penalty $\eta > 0$. This proximity operator is easy to compute since $\| {\bf{x}} \|_q^q$ is separable and the computation of ${\rm{pro}}{{\rm{x}}_{q,\eta }}$ reduces to solving a number of univariate minimization problems.

When $q=0$, the solution is explicitly given by
\begin{equation} % (7)
\mathrm{prox}_{0,\eta}(\mathbf{t})_i = {H_\eta }{({\bf{t}})_i}=\left\{
\begin{aligned}
&0,~~~~~~~|t_i|<\sqrt{2/\eta} \\
&\{0,t_i\},~|t_i|=\sqrt{2/\eta}\\
&t_i,~~~~~~~\mathrm{otherwise}
\end{aligned}
\right.
\end{equation}
for $i=1,\cdots,m$, which is the well-known hard-thresholding operation. When $q=1$, this proximity operator is the well-known soft-thresholding or shrinkage operator and has a closed-form expression as
\begin{equation} % (8)
{\rm{pro}}{{\rm{x}}_{1,\eta }}{({\bf{t}})_i} = {S_\eta }{({\bf{t}})_i} = {\rm{sign}}({t_i})\max \left\{ {{\rm{|}}{t_i}{\rm{|}} - {1 \mathord{\left/
 {\vphantom {1 \eta }} \right.
 \kern-\nulldelimiterspace} \eta },0} \right\}
\end{equation}
for $i = 1, \cdots ,m$.

When $0 < q < 1$, it can be computed as [37]
\begin{equation} % (9)
\mathrm{prox}_{q,\eta}(\mathbf{t})_i =\left\{
\begin{aligned}
&0,~~~~~~~~~~~~~~~~|t_i|< \tau \\
&{\{ 0,{\rm{sign}}({t_i})\beta \} },~|t_i|= \tau\\
&\mathrm{sign}(t_i)z_i,~~~~~~|t_i|> \tau
\end{aligned}
\right.,~i=1,\cdots,m
\end{equation}
where $\beta  = {[2(1 - q)/\eta ]^{\frac{1}{{2 - q}}}}$, $\tau  = \beta  + q{\beta ^{q - 1}}/\eta$, $z_i$ is the solution of $h(z) = q{z^{q - 1}} + \eta z - \eta \left| {{t_i}} \right| = 0$ over the region $(\beta ,{\rm{|}}{t_i}{\rm{|}})$. Since $h(z)$ is convex, when ${\rm{|}}{t_i}{\rm{|}} > \tau$, ${z_i}$ can be efficiently solved using a Newton's method. For the special cases of $q = 1/2$ or $q = 2/3$, the proximal mapping can be explicitly expressed as the solution of a cubic or quartic equation [38].

\section{Proposed Algorithms}

Generally, the linearly constrained ${\ell _{{q_1}}}-{\ell _{{q_2}}}$ mixed minimization problem (2) is difficult to tackle since both terms in the objective are nonconvex and nonsmooth.
It can be directly solved by the standard two-block ADMM procedure, but it is not guaranteed to converge in the nonconvex case of $0 \le {q_{\rm{1}}},{q_2} < 1$.
Empirical studies show that the directly extended two-blocks ADMM algorithm for (2) often fails to converge (see Fig. 1 in section V).
To address this problem, we propose to solve (2) approximately and develop two first-order algorithms, which are guaranteed to converge in the nonconvex case.
The first algorithm is based on the proximal BCD and the second one is a four-block ADMM algorithm.

First, we consider an approximation of (2) as
\begin{equation}   %%(10)
\begin{split}
&\mathop {\min }\limits_{{{\bf{x}}_1},{{\bf{x}}_2}} \left\{ {\mu \left\| {{{\bf{x}}_1}} \right\|_{{q_1}}^{{q_1}} + \left\| {{{\bf{x}}_2}} \right\|_{{q_2}}^{{q_2}}} \right\}\\
&\mathrm{subject~to}~~
{\left\| {{{\bf{A}}_1}{{\bf{x}}_1} + {{\bf{A}}_2}{{\bf{x}}_2} - {\bf{y}}} \right\|_2} \le \varepsilon
\end{split}
\end{equation}
where $\varepsilon  > 0$. It is easy to see that as $\varepsilon  \to 0$, the problem (10) reduces to the problem (2). Thus, with a sufficient small $\varepsilon$, the solution of (10) accurately approaches that of (2). Further, this constrained optimization problem can be converted into an alternative unconstrained form
\begin{equation}   %%(11)
\mathop {\min }\limits_{{{\bf{x}}_{\rm{1}}},{{\bf{x}}_2}} \left\{ {\frac{1}{\beta }\left\| {{{\bf{A}}_1}{{\bf{x}}_1} + {{\bf{A}}_2}{{\bf{x}}_2} - {\bf{y}}} \right\|_2^2 + \mu \left\| {{{\bf{x}}_1}} \right\|_{{q_1}}^{{q_1}} + \left\| {{{\bf{x}}_2}} \right\|_{{q_2}}^{{q_2}}} \right\}
\end{equation}
where $\beta  > 0$ is a penalty parameter. A small $\varepsilon $ in (10) corresponds to a small $\beta$ in the problem (11).  As $\beta  \to 0$, the solutions of (11) satisfy ${\left\| {{{\bf{A}}_1}{{\bf{x}}_1} + {{\bf{A}}_2}{{\bf{x}}_2} - {\bf{y}}} \right\|_2} \to 0$ and the problem (11) reduces to the problem (2). Thus, we can use a sufficient small $\beta$ to enforce ${\left\| {{{\bf{A}}_1}{{\bf{x}}_1} + {{\bf{A}}_2}{{\bf{x}}_2} - {\bf{y}}} \right\|_2} \approx 0$, e.g., $\beta  = {10^{ - 6}}$ in the experiments in section V.

Note that, although the formulation (11) is an approximation of (2), it is a more reasonable formulation in some applications where the measurements contains additive Gaussian noise. Specifically, in the presence of measurement noise, the signal model becomes
\begin{equation}   %%(12)
{\bf{y}} = {{\bf{A}}_{\rm{1}}}{{\bf{x}}_{\rm{1}}}{\rm{ + }}{{\bf{A}}_{\rm{2}}}{{\bf{x}}_{\rm{2}}} + {\bf{n}}
\end{equation}
where ${\bf{n}}$ is the noise. In this case, the formulations (10) and (11) are more reasonable than (2) as they take the measurement noise into account.
In the following, we develop two algorithms for (11) based on the BCD and ADMM frameworks, respectively.

\subsection{Proximal BCD Algorithm}

The core idea of the BCD algorithm is to solve an intractable optimization problem by successively performing approximate minimization along coordinate directions or coordinate hyperplanes. Specifically, for the problem (11), at the $k+1$-th iteration, ${{\bf{x}}_1}$ and ${{\bf{x}}_2}$ are alternatingly updated by minimizing the objective as
\begin{align}   %%(13)(14)
&{\bf{x}}_1^{k + 1} = \arg\mathop {\min }\limits_{{{\bf{x}}_{\rm{1}}}} \left\{ {\frac{1}{\beta }\left\| {{{\bf{A}}_1}{{\bf{x}}_1} + {{\bf{A}}_2}{\bf{x}}_2^k - {\bf{y}}} \right\|_2^2 + \mu \left\| {{{\bf{x}}_1}} \right\|_{{q_1}}^{{q_1}}} \right\}\\
&{\bf{x}}_2^{k + 1} = \arg\mathop {\min }\limits_{{{\bf{x}}_2}} \left\{ {\frac{1}{\beta }\left\| {{{\bf{A}}_1}{\bf{x}}_1^{k + 1} + {{\bf{A}}_2}{{\bf{x}}_2} - {\bf{y}}} \right\|_2^2 + \left\| {{{\bf{x}}_2}} \right\|_{{q_2}}^{{q_2}}} \right\}.
\end{align}
Since it is difficult to exactly minimize these two nonconvex and nonsmooth subproblems, a standard trick is to adopt an approximation of this scheme via the proximal linearization of each subproblem. Specifically, consider a quadratic majorization of the first term in (13) as
\begin{equation}
\begin{split}
&\left\| {{{\bf{A}}_1}{{\bf{x}}_1} + {{\bf{A}}_2}{\bf{x}}_2^k - {\bf{y}}} \right\|_2^2 \approx \left\| {{{\bf{A}}_1}{\bf{x}}_1^k + {{\bf{A}}_2}{\bf{x}}_2^k - {\bf{y}}} \right\|_2^2 \\
&~~~~~~~~~~~~~~~~~~~~~+ \left\langle {{{\bf{x}}_1} - {\bf{x}}_1^k,{g_1}({\bf{x}}_1^k)} \right\rangle  + \frac{{{\eta _1}}}{2}\left\| {{{\bf{x}}_1} - {\bf{x}}_1^k} \right\|_2^2 \notag
\end{split}
\end{equation}
where ${g_1}({\bf{x}}_1^k) = 2{\bf{A}}_1^T({{\bf{A}}_1}{\bf{x}}_1^k + {{\bf{A}}_2}{\bf{x}}_2^k - {\bf{y}})$, ${\eta _1} > 0$ is a proximal parameter. With this approximation, the ${{\bf{x}}_1}$-update step becomes a form of the proximity operator (6), which can be efficiently updated as
\begin{equation}   %%(15)
{\bf{x}}_1^{k + 1}= {\rm{pro}}{{\rm{x}}_{{q_1},{\eta _1}/(\beta \mu )}}({\bf{c}}_1^k) = \left\{
 \begin{aligned}
&{H_{{\eta _1}/(\beta \mu )}}({\bf{c}}_1^k),~{{q_1} = 0}\\
&{\rm{solved~as~(9)}},~{0 < {q_1} < 1}\\
&{S_{{\eta _1}/(\beta \mu )}}({\bf{c}}_1^k),~~{{q_1} = 1}
\end{aligned}
\right.
\end{equation}
where ${\bf{c}}_1^k = {\bf{x}}_1^k - \frac{2}{{{\eta _1}}}{\bf{A}}_1^T({{\bf{A}}_1}{\bf{x}}_1^k + {{\bf{A}}_2}{\bf{x}}_2^k - {\bf{y}})$. In a similar manner, we use a quadratic majorization of the first term in (14) with a proximal parameter ${\eta _2} > 0$. Then, the ${{\bf{x}}_2}$-update step (14) can be solved as
\begin{equation}   %%(16)
{\bf{x}}_2^{k + 1}= {\rm{pro}}{{\rm{x}}_{{q_2},{\eta _2}/\beta}}({\bf{c}}_2^k) = \left\{
 \begin{aligned}
&{H_{{\eta _2}/\beta}}({\bf{c}}_2^k),~~~~{{q_2} = 0}\\
&{\rm{solved~as~(9)}},~{0 < {q_2} < 1}\\
&{S_{{\eta _2}/\beta}}({\bf{c}}_1^k),~~~~~{{q_2} = 1}
\end{aligned}
\right.
\end{equation}
where ${\bf{c}}_2^k = {\bf{x}}_2^k - \frac{2}{{{\eta _2}}}{\bf{A}}_2^T({{\bf{A}}_1}{\bf{x}}_1^{k + 1} + {{\bf{A}}_2}{\bf{x}}_2^k - {\bf{y}})$.

This algorithm can scale to relatively large problems since the dominant computational complexity in each iteration is the cheap matrix-vector multiplication.
The convergence condition for this kind of nonconvex BCD algorithm has been established recently in [39]. As shown in the following result, under some mild conditions, the above two-block coodinate descent procedure is guaranteed to be globally convergent in the nonconvex case.

\textbf{Theorem 1 ([39]).} For any ${q_1} \ge 0$ and ${q_2} \ge 0$, if ${\eta _1} > 2{\lambda _{\max }}({\bf{A}}_1^T{{\bf{A}}_1})$ and ${\eta _2} > 2{\lambda _{\max }}({\bf{A}}_2^T{{\bf{A}}_2})$, the algorithm updated via (15) and (16) is a descent algorithm and the generated sequence $\{ ({\bf{x}}_1^k,{\bf{x}}_2^k)\}$ converges to a critical point of the problem (11).

\subsection{ADMM Algorithm}

ADMM is a powerful framework which is well suited to solve many high-dimensional optimization problems [36]. ADMM uses a decomposition-coordination procedure to naturally decouple the variables, which makes the global problem easy to tackle. Specifically, using two auxiliary variables ${{\bf{z}}_1} = {{\bf{x}}_1}$ and ${{\bf{z}}_2} = {{\bf{x}}_2}$, (11) can be equivalently reformulated as
\begin{align}%%(17)
\nonumber
&\mathop {\min }\limits_{{{\bf{x}}_{\rm{1}}},{{\bf{x}}_2},{{\bf{z}}_{\rm{1}}},{{\bf{z}}_2}} \left\{ {\left\| {{{\bf{A}}_1}{{\bf{x}}_1} + {{\bf{A}}_2}{{\bf{x}}_2} - {\bf{y}}} \right\|_2^2 + \beta \mu \left\| {{{\bf{z}}_1}} \right\|_{{q_1}}^{{q_1}} + \beta \left\| {{{\bf{z}}_2}} \right\|_{{q_2}}^{{q_2}}} \right\}\\
&~~~~~~~~~~~~~~~~~~~~~~~~~\mathrm{subject~to}~~ {{\bf{x}}_1} = {{\bf{z}}_1},~{{\bf{x}}_2} = {{\bf{z}}_2}.
\end{align}
The augmented Lagrangian function is
\begin{equation}
\begin{split}
&\mathcal{L}({{\bf{x}}_1},{{\bf{x}}_2},{{\bf{z}}_1},{{\bf{z}}_2},{{\bf{w}}_1},{{\bf{w}}_2}) = \left\| {{{\bf{A}}_1}{{\bf{x}}_1} + {{\bf{A}}_2}{{\bf{x}}_2} - {\bf{y}}} \right\|_2^2\\
& + \beta \mu \left\| {{{\bf{z}}_1}} \right\|_{{q_1}}^{{q_1}} + \beta \left\| {{{\bf{z}}_2}} \right\|_{{q_2}}^{{q_2}} + \left\langle {{{\bf{w}}_1},{{\bf{x}}_1} - {{\bf{z}}_1}} \right\rangle  + \left\langle {{{\bf{w}}_2},{{\bf{x}}_2} - {{\bf{z}}_2}} \right\rangle\\
& + \frac{{{\rho _1}}}{2}\left\| {{{\bf{x}}_1} - {{\bf{z}}_1}} \right\|_2^2 + \frac{{{\rho _2}}}{2}\left\| {{{\bf{x}}_2} - {{\bf{z}}_2}} \right\|_2^2 \notag
\end{split}
\end{equation}
where ${{\bf{w}}_1}$ and ${{\bf{w}}_2}$ are the dual variables, ${\rho _1}$ and ${\rho _2}$ are positive penalty parameters. ADMM iteratively updates the primal and dual variables as follows
\begin{align}
{\bf{z}}_{\rm{1}}^{k + 1} = \arg \mathop {\min }\limits_{{{\bf{z}}_1}} \bigg( {\beta \mu \left\| {{{\bf{z}}_1}} \right\|_{{q_1}}^{{q_1}} + \frac{{{\rho _1}}}{2}\left\| {{\bf{x}}_1^k - {{\bf{z}}_1} \!+\! \frac{{\bf{w}}_1^k}{\rho _1}} \right\|_2^2} \bigg)
\end{align}
\begin{align}
{\bf{z}}_2^{k + 1} = \arg \mathop {\min }\limits_{{{\bf{z}}_2}} \bigg( {\beta \left\| {{{\bf{z}}_2}} \right\|_{{q_2}}^{{q_2}} + \frac{{{\rho _2}}}{2}\left\| {{\bf{x}}_2^k - {{\bf{z}}_2} + \frac{{\bf{w}}_2^k}{\rho _2}} \right\|_2^2} \bigg)~
\end{align}
\begin{align}
\nonumber
{\bf{x}}_1^{k + 1} &= \arg \mathop {\min }\limits_{{{\bf{x}}_1}} \bigg( \left\| {{{\bf{A}}_1}{{\bf{x}}_1} + {{\bf{A}}_2}{\bf{x}}_2^k - {\bf{y}}} \right\|_2^2\\
 &~~~~~~~~~~~~~~~~~~~~+ \frac{{{\rho _1}}}{2}\left\| {{{\bf{x}}_1} - {\bf{z}}_1^{k + 1} + \frac{{\bf{w}}_1^k}{\rho _1}} \right\|_2^2 \bigg)
\end{align}
\begin{align}
\nonumber
{\bf{x}}_2^{k + 1} &= \arg \mathop {\min }\limits_{{{\bf{x}}_2}} \bigg( \left\| {{{\bf{A}}_1}{\bf{x}}_1^{k + 1} + {{\bf{A}}_2}{{\bf{x}}_2} - {\bf{y}}} \right\|_2^2\\
 &~~~~~~~~~~~~~~~~~~~~+ \frac{{{\rho _2}}}{2}\left\| {{{\bf{x}}_2} - {\bf{z}}_2^{k + 1} + \frac{{\bf{w}}_2^k}{\rho_2}} \right\|_2^2 \bigg)
\end{align}
\begin{align}
{\bf{w}}_1^{k + 1} &= {\bf{w}}_1^k + {\rho _1}({\bf{x}}_1^{k + 1} - {\bf{z}}_1^{k + 1})~
\end{align}
\begin{align}
{\bf{w}}_2^{k + 1} &= {\bf{w}}_2^k + {\rho _2}({\bf{x}}_2^{k + 1} - {\bf{z}}_2^{k + 1}).
\end{align}
Both the ${{\bf{z}}_1}$- and ${{\bf{z}}_2}$-subproblems are the form of the proximity operator (6) and can be updated as (7), (8) and (9). The objective function in the ${{\bf{x}}_1}$- and ${{\bf{x}}_2}$-subproblems are quadratic, the exact solutions are directly given by
\begin{align} %(24) (25)
\nonumber
{\bf{x}}_1^{k + 1} &= {(2{\bf{A}}_1^T{{\bf{A}}_1} + {\rho _1}{\bf{I}})^{ - 1}}[ 2{\bf{A}}_1^T({\bf{y}} - {{\bf{A}}_2}{\bf{x}}_2^k)\\
&~~~~~~~~~~~~~~~~~~~~~~~~~~~~~~~~~~~~~ + {\rho _1}{\bf{z}}_1^{k + 1} - {\bf{w}}_1^k ]\\ \nonumber
{\bf{x}}_2^{k + 1} &= {(2{\bf{A}}_2^T{{\bf{A}}_2} + {\rho _2}{\bf{I}})^{ - 1}}[ 2{\bf{A}}_2^T({\bf{y}} - {{\bf{A}}_1}{\bf{x}}_1^{k + 1})\\
&~~~~~~~~~~~~~~~~~~~~~~~~~~~~~~~~~~~~~ + {\rho _2}{\bf{z}}_2^{k + 1} - {\bf{w}}_2^k ].
\end{align}
In computing the inverse in (24) and (25), Cholesky decomposition can be used to reduce the computational complexity [36]. When the penalty parameters ${\rho_1}$ and ${\rho_2}$ do not change in iteration, we can only compute the inverse  once.
Moreover, when ${{\bf{A}}_i}$ is orthonormal, i.e., ${{\bf{A}}_i}{\bf{A}}_i^T = {\bf{I}}$, the inversion in the ${{\bf{x}}_i}$-step can be avoided as
\begin{equation} %(26)
{(2{\bf{A}}_i^T{{\bf{A}}_i} + {\rho _i}{\bf{I}})^{ - 1}} = \frac{1}{{{\rho _i}}}{\bf{I}} - \frac{2}{{{\rho _i}(2 + {\rho _i})}}{\bf{A}}_i^T{{\bf{A}}_i}.\notag
\end{equation}

In the following, we provide a sufficient condition for the convergence of the above ADMM algorithm.

\textbf{Theorem 2.} Let ${\lambda _i} = {\lambda _{\max }}({\bf{A}}_i^T{{\bf{A}}_i})$ and ${\varphi _i} = {\lambda _{\min }}({\bf{A}}_i^T{{\bf{A}}_i})$, $i = 1,2$, for any ${q_1} \ge 0$ and ${q_2} \ge 0$, if
\begin{equation} %(26)
\begin{split}
{\rho _1} &> \frac{{16\lambda _1^2}}{{{\rho _1}}} + \frac{{16{\lambda _1}{\lambda _2}}}{{{\rho _2}}} - 2{\varphi _1},\\
{\rho _2} &> \frac{{16\lambda _2^2}}{{{\rho _2}}} + \frac{{16{\lambda _1}{\lambda _2}}}{{{\rho _1}}} - 2{\varphi _2},
\end{split}
\end{equation}
the sequence $\{ ({\bf{z}}_1^k,{\bf{z}}_2^k,{\bf{x}}_1^k,{\bf{x}}_2^k,{\bf{w}}_1^k,{\bf{w}}_2^k)\}$ generated by the ADMM algorithm via (18)--(23) converges to a critical point of the problem (11).

\textit{Proof:} See Appendix A.

The convergence properties of ADMM for the nonconvex case have been established very recently in [40], [41]. This convergence condition for the above 4-block ADMM algorithm is derived via extending the result for 2-block ADMM in [40]. It is worth stressing that, there exists a recent work [46] on the convergence of nonconvex multi-block ADMM. However, the convergence condition in Theorem 2 cannot be directly derived from the results in [46],
since [46] only considers the class of ADMM algorithms with a single dual variable while our algorithm has multiple (two) dual variables.

\section{Multichannel Joint Recovery for Color Images}

In recovering a color image with 3 channels (e.g., RGB image), the above BCD and ADMM algorithms can be used to recover each channel independently. However, since the original 3 channel images (also the corruption in the three channels) may have similar sparsity pattern, performance improvement can be expected via exploiting the feature correlation among different channels, also called group or joint sparsity in multitask sparse recovery. In this section, we extend the above BCD and ADMM algorithms to the multitask case.

In the multitask case, the linear measurements ${\bf{Y}} \in {\mathbb{R}^{m \times L}}$ of $L$ channels can be modeled as
\begin{equation} %(27)
{\bf{Y}} = {{\bf{A}}_{\rm{1}}}{{\bf{X}}_{\rm{1}}}{\rm{ + }}{{\bf{A}}_{\rm{2}}}{{\bf{X}}_{\rm{2}}}
\end{equation}
where ${{\bf{X}}_k} \in {\mathbb{R}^{{n_k} \times L}}$, $k = 1,2$, are the sparse features in the two components. To exploit the joint sparsity among the $L$ channels, we consider a multitask version of the problem (11) as
\begin{equation} %(28)
\mathop {\min }\limits_{{{\bf{X}}_{\rm{1}}},{{\bf{X}}_2}} \bigg\{ {\frac{1}{\beta }\left\| {{{\bf{A}}_1}{{\bf{X}}_1} \!+\! {{\bf{A}}_2}{{\bf{X}}_2} \!-\! {\bf{Y}}} \right\|_F^2 + \mu \left\| {{{\bf{X}}_1}} \right\|_{2,{q_1}}^{{q_1}} + \left\| {{{\bf{X}}_2}} \right\|_{2,{q_2}}^{{q_2}}} \bigg\}
\end{equation}
where $0 \le {q_{\rm{1}}},{q_2} < 1$, $\left\| {\bf{X}} \right\|_{2,q}^q$ is defined as
\begin{equation}
\left\| {\bf{X}} \right\|_{2,q}^q = \sum\nolimits_i {\left\| {{\bf{X}}[i,:]} \right\|_2^q}  = \sum\nolimits_i {{{\left( {\sum\nolimits_j {{{\bf{X}}^2}[i,j]} } \right)}^{q/2}}} .\notag
\end{equation}

Note that, in other joint sparse recovery applications, such as multiple measurement vectors recovery in CS, the formulation (28) can be modified to enforce joint sparsity only on one of the features.

Before presenting the algorithms, we give a generalization of the $\ell_q$-norm proximity operator.

\textbf{Theorem 3.} For any $0 \le q \le 1$, $\eta  > 0$, ${\bf{x}} \in {\mathbb{R}^L}$, consider the following vector optimization problem
\begin{equation} %(29)
\mathop {\min }\limits_{\bf{x}} \left\{ {\left\| {\bf{x}} \right\|_{2}^q + \frac{\eta }{2}\left\| {{\bf{x}} - {\bf{t}}} \right\|_2^2} \right\}.
\end{equation}
Then, its solution is given by
\begin{equation} %(30)
{\bf{x}}{\rm{ = pro}}{{\rm{x}}_{q,\eta \|{\bf{t}}\|_2^{2 - q}}}(1) \cdot {\bf{t}}.
\end{equation}

\textit{Proof:} See Appendix B. For the special case of $q=1$, (29) reduces to the $\ell_1$-norm proximity operator of multi-task which has been addressed in [47].

\subsection{BCD Algorithm for Multitask}
Using a similar linearization strategy as in the BCD algorithm for signle-task in section III, the BCD algorithm for the multitask problem (28) consists of the following two steps
\begin{equation} %(31)
\begin{split}
&{\bf{X}}_1^{k + 1}= \arg\mathop {\min }\limits_{{{\bf{X}}_1}} \bigg\{\mu \left\| {{{\bf{X}}_1}} \right\|_{2,{q_1}}^{{q_1}}\\
 &+ {\frac{{{\eta _3}}}{{2\beta }}\Big\| {{{\bf{X}}_1} - {\bf{X}}_1^k + \frac{2}{{{\eta _3}}}{\bf{A}}_1^T({{\bf{A}}_1}{\bf{X}}_1^k + {{\bf{A}}_2}{\bf{X}}_2^k \!-\! {\bf{Y}})} \Big\|_F^2} \bigg\}
\end{split}
\end{equation}
\begin{equation} %(32)
\begin{split}
&{\bf{X}}_2^{k + 1} = \arg\mathop {\min }\limits_{{{\bf{X}}_2}} \bigg\{\left\| {{{\bf{X}}_2}} \right\|_{2,{q_1}}^{{q_1}}\\
&+{\frac{{{\eta _4}}}{{2\beta }}\Big\| {{{\bf{X}}_2} - {\bf{X}}_2^k \!+\! \frac{2}{{{\eta _4}}}{\bf{A}}_2^T({{\bf{A}}_1}{\bf{X}}_1^{k + 1} \!+\! {{\bf{A}}_2}{\bf{X}}_2^k \!-\! {\bf{Y}})} \Big\|_F^2} \bigg\}
\end{split}
\end{equation}
where ${\eta _3} > 0$ and ${\eta _4} > 0$ are proximal parameters used in the linearization. These two subproblems can be solved row-wise as (29). The following sufficient condition for the convergence of this algorithm can be derived following similarly to the work [39].

\textbf{Theorem 4.} For any ${q_1} \ge 0$ and ${q_2} \ge 0$, if ${\eta _3} > 2{\lambda _{\max }}({\bf{A}}_1^T{{\bf{A}}_1})$ and ${\eta _4} > 2{\lambda _{\max }}({\bf{A}}_2^T{{\bf{A}}_2})$, the algorithm updated via (31) and (32) is a descent algorithm and the generated sequence $\{ ({\bf{X}}_1^k,{\bf{X}}_2^k)\}$ converges to a critical point of the problem (28).

\subsection{ADMM Algorithm for Multitask}

Using two auxiliary variables ${{\bf{Z}}_1} = {{\bf{X}}_1}$ and ${{\bf{Z}}_2} = {{\bf{X}}_2}$, the problem (28) can be equivalently reformulated as
\begin{align}%%(33)
\nonumber
&\mathop {\min }\limits_{{{\bf{X}}_{\rm{1}}},{{\bf{X}}_2},{{\bf{Z}}_{\rm{1}}},{{\bf{Z}}_2}} \Big\{ \left\| {{{\bf{A}}_1}{{\bf{X}}_1} + {{\bf{A}}_2}{{\bf{X}}_2} - {\bf{Y}}} \right\|_F^2\\ \nonumber
&~~~~~~~~~~~~~~~~~~~~~~~~~~~~ + \beta \mu \left\| {{{\bf{Z}}_1}} \right\|_{2,{q_1}}^{{q_1}} + \beta \left\| {{{\bf{Z}}_2}} \right\|_{2,{q_2}}^{{q_2}} \Big\}\\
&~~~~~~~~~~~~~~~~~~~~~~~~~\mathrm{subject~to}~~ {{\bf{Z}}_1} = {{\bf{X}}_1},~{{\bf{Z}}_2} = {{\bf{X}}_2}.
\end{align}

Then, similar to the ADMM algorithm in section III, the ADMM algorithm for the multitask problem (28) consists of the following steps
\begin{align}
{\bf{Z}}_1^{k + 1}= &\arg \mathop {\min }\limits_{{{\bf{Z}}_1}} \left( {\beta \mu \left\| {{{\bf{Z}}_1}} \right\|_{2,{q_1}}^{{q_1}} + \frac{{{\rho _3}}}{2}\left\| {{\bf{X}}_1^k - {{\bf{Z}}_1} + \frac{{{\bf{W}}_1^k}}{{{\rho _3}}}} \right\|_F^2} \right)\\
{\bf{Z}}_2^{k + 1}= &\arg \mathop {\min }\limits_{{{\bf{Z}}_2}} \left( {\beta \left\| {\bf{Z}} \right\|_{2,{q_2}}^{{q_2}} + \frac{{{\rho _4}}}{2}\left\| {{\bf{X}}_2^k - {{\bf{Z}}_2} + \frac{{{\bf{W}}_2^k}}{{{\rho _4}}}} \right\|_F^2} \right)\\ \nonumber
{\bf{X}}_1^{k + 1}= &{(2{\bf{A}}_1^T{{\bf{A}}_1} + {\rho_3}{\bf{I}})^{-1}}\big[ 2{\bf{A}}_1^T({\bf{Y}} - {{\bf{A}}_2}{\bf{X}}_2^k)\\
&~~~~~~~~~~~~~~~~~~~~~~~~~~~~~~~ + {\rho _3}{\bf{Z}}_1^{k + 1} - {\bf{W}}_1^k \big]\\ \nonumber
{\bf{X}}_2^{k + 1}= &{(2{\bf{A}}_2^T{{\bf{A}}_2} + {\rho_4}{\bf{I}})^{-1}}\big[ 2{\bf{A}}_2^T({\bf{Y}} - {{\bf{A}}_1}{\bf{X}}_1^{k + 1})\\
&~~~~~~~~~~~~~~~~~~~~~~~~~~~~~~~ + {\rho _4}{\bf{Z}}_2^{k + 1} - {\bf{W}}_2^k \big]\\
{\bf{W}}_1^{k + 1} &= {\bf{W}}_1^k + {\rho _3}({\bf{X}}_1^{k + 1} - {\bf{Z}}_1^{k + 1})\\
{\bf{W}}_2^{k + 1} &= {\bf{W}}_2^k + {\rho _4}({\bf{X}}_2^{k + 1} - {\bf{Z}}_2^{k + 1}).
\end{align}
${{\bf{W}}_1}$ and ${{\bf{W}}_2}$ are the dual variables, ${\rho _3} > 0$ and ${\rho _4} > 0$ are penalty parameters. The ${{\bf{Z}}_1}$- and ${{\bf{Z}}_2}$-subproblems can be solved row-wise as (29). The following sufficient condition for the convergence of this ADMM algorihm can be derived similarly to Theorem 2.

\textbf{Theorem 5.} Let ${\lambda _i} = {\lambda _{\max }}({\bf{A}}_i^T{{\bf{A}}_i})$ and ${\varphi _i} = {\lambda _{\min }}({\bf{A}}_i^T{{\bf{A}}_i})$, $i = 1,2$, for any ${q_1} \ge 0$ and ${q_2} \ge 0$, if
\begin{equation}
\begin{split}
{\rho _3} &> \frac{{16\lambda _1^2}}{{{\rho _3}}} + \frac{{16{\lambda _1}{\lambda _2}}}{{{\rho _4}}} - 2{\varphi _1},\\
{\rho _4} &> \frac{{16\lambda _2^2}}{{{\rho _4}}} + \frac{{16{\lambda _1}{\lambda _2}}}{{{\rho _3}}} - 2{\varphi _2},\notag
\end{split}
\end{equation}
the sequence $\{ ({\bf{Z}}_1^k,{\bf{Z}}_2^k,{\bf{X}}_1^k,{\bf{X}}_2^k,{\bf{W}}_1^k,{\bf{W}}_2^k)\}$ generated by the ADMM algorithm via (34)--(39) converges to a critical point of the problem (33).

When $L = 1$, these two algorithms reduces to the BCD and ADMM algorithms for single task in section III.

\section{Numerical Experiments}

In this section, we evaluate the performance of the new methods via three groups of experiments, including a synthetic sparse separation experiment, inpainting experiments, and an experiment of robust compressive sensing in impulsive noise.

Selecting an appropriate value of $\mu$ is important for the new algorithms (as well as the compared FISTA and YALL1 methods) to achieve satisfactory performance.
In general, the optimal value is related with the statistical information of the true signal components, the values of ${q_1}$ and ${q_2}$, and hence is difficult to obtain. Various suboptimal approaches can be used for the selection. For example, it can be selected based on experience or by learning. Typically, using training data, it can be learned via cross validation [44]. Another popular approach is to compute the restoration for a set of $\mu$, which is often called the regularization path, and select the optimal value based on some statistical information of ${{\bf{x}}_2}$. In practice, for each algorithm the optimal value of $\mu$ could be different. To compare the algorithms fairly, in each algorithm $\mu$ is chosen by providing the best performance, in terms of the lowest relative error (RelErr) of recovery.

In the proposed BCD and ADMM algorithms, we use $\beta  = {10^{ - 6}}$. Generally, with a very small value of $\beta$, both the algorithms would be very slow and impractical. A standard trick to accelerate the algorithms is to adopt a continuation process for this parameter, e.g., use a properly large starting value of it and gradually decrease it by iteration until reaching the target value, e.g., ${\beta _0} \ge {\beta _1} \ge  \cdots  \ge {\beta _K} = {\beta _{K + 1}} =  \cdots  = \beta$. In the implementation, we use a continuation process for $\beta$ as ${\beta _k} = 0.97{\beta _{k - 1}}$ if ${\beta _k} > {10^{ - 6}}$ and ${\beta _k} = {10^{ - 6}}$ otherwise.

\subsection{Synthetic Experiment for Sparse Signals Separation}

We first evaluate the new algorithms by a synthetic experiment with ${{\bf{A}}_1} \in {\mathbb{R}^{128 \times 128}}$ be a DCT matrix and ${{\bf{A}}_2} \in {\mathbb{R}^{128 \times 128}}$ be an orthonormal Gaussian random matrix. ${{\bf{x}}_1}$ and ${{\bf{x}}_2}$ have the same sparsity of $K$. The positions of the $K$ nonzeros are uniformly randomly chosen while the amplitude of each nonzero entry follows a Gaussian distribution.

Fig. 1 shows the typical convergence behavior of the proposed BCD and ADMM algorithms in nonconvex conditions, in comparison with the standard ADMM (S-ADMM) algorithm applied to (2) (see Appendix C). For S-ADMM, we use $\rho = 10$, ${c_i} = 2.1{\lambda _{\max }}({\bf{A}}_i^T{{\bf{A}}_i})$, $i = 1,2$. With this setting, the corresponding Lagrangian function is guaranteed to decrease in both the ${{\bf{x}}_1}$- and ${{\bf{x}}_2}$-steps even in the nonconvex case. It can be seen that S-ADMM does not converge in the nonconvex cases.

Fig. 2 compares the performance of the algorithms versus $K$ in terms of success rate of recovery. A recovery ${\hat{\bf{ x}}_1}$ is regarded as successful if the RelErr satisfies $\frac{\left\| {{{\hat{\bf{ x}}}_1} - {{\bf{x}}_1}} \right\|_2}{\left\| {{{\bf{x}}_1}} \right\|_2} \le {10^{ - 2}}$. The result is an average over 300 independent runs. The S-ADMM algorithm with $q_{1} = {q_2} = 1$ and the FISTA algorithm [48] solving (11) are included for comparison. With ${q_{\rm{1}}} = {q_2} = 1$, S-ADMM is guaranteed to globally converge [36]. In the nonconvex case of ${q_1} < 1$ and/or ${q_2} < 1$, while the proposed BCD and ADMM algorithms are guaranteed to converge under some mild conditions, there is no guarantee of convergence for FISTA. In the nonconvex case of ${q_1} < 1$ and/or ${q_2} < 1$, FISTA and the proposed algorithms are initialized by S-ADMM with $q_{1} = {q_2} = 1$ and $\mu=1$. Fig. 3 presents the recovery performance of the FISTA, BCD and ADMM methods for different values of $q_1$ and $q_2$.

It can be seen that, with ${q_1} < 1$ and ${q_2} < 1$, each of the three nonconvex methods can significantly outperform the convex S-ADMM method (with $q_{1} = {q_2} = 1$). Note that, when $q_{1} = {q_2} = 1$, all the FISTA, BCD and ADMM methods can find a global minimizer of (11) and achieve the same accuracy (which approximates the accuracy of S-ADMM with $q_{1} = {q_2} = 1$ since $\beta  = {10^{ - 6}}$ is very small). The results indicate that relatively small values of $q_1$ and $q_2$ (e.g., ${q_{1}},{q_2} < 0.5$) tend to yield better performance.

\begin{figure}[!t]
 \centering
 \includegraphics[scale = 0.65]{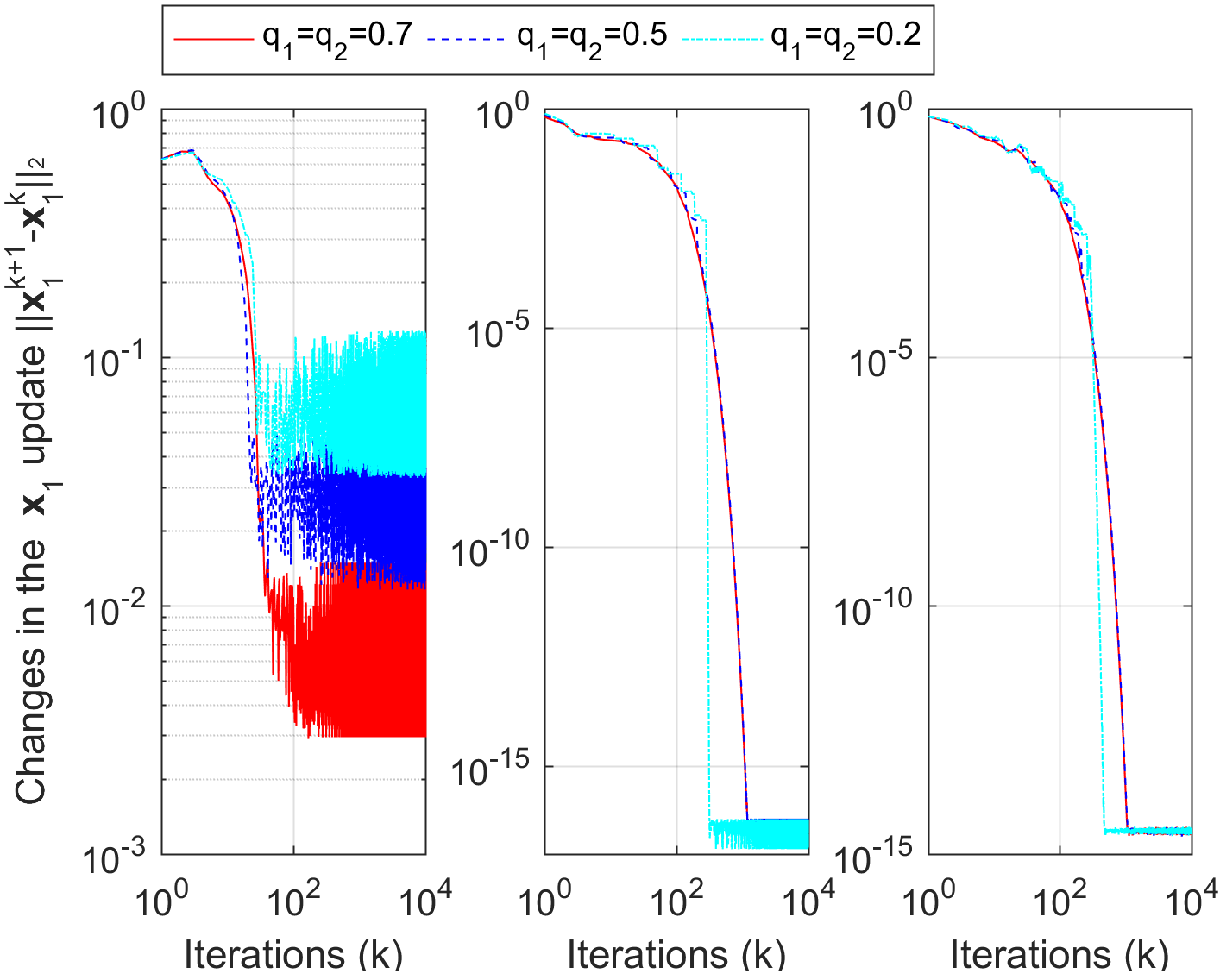}
\caption{Typical convergence behavior in the nonconvex case, with $\mu=1$ and $K=20$. \textit{Left}: S-ADMM. \textit{Middle}: BCD. \textit{Right}: ADMM.}
 \label{figure1}
\end{figure}

\begin{figure}[!t]
 \centering
 \includegraphics[scale = 0.20]{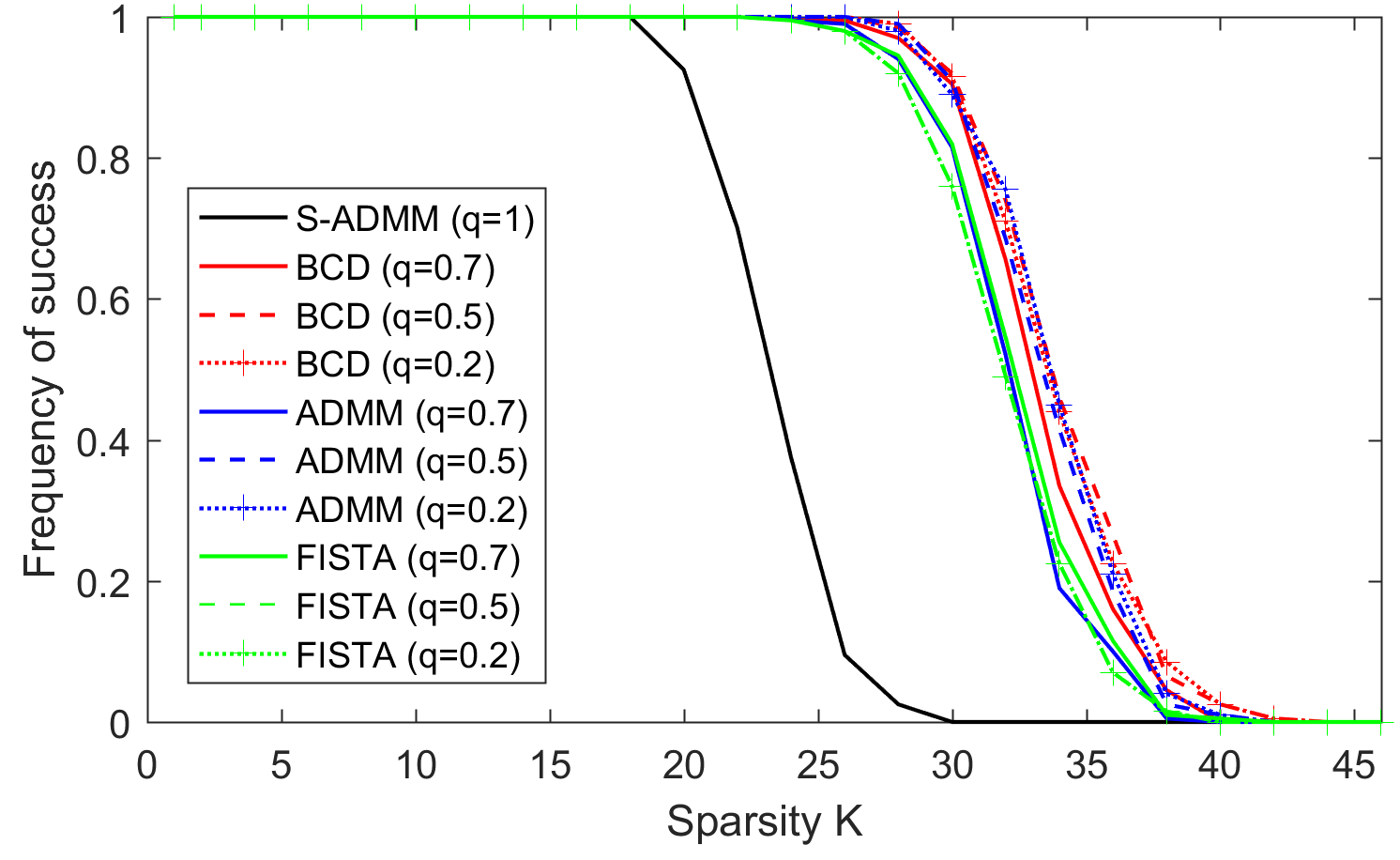}
\caption{Frequency of successful recovery versus sparsity, ${\bf{A}}_1$ is a DCT matrix, ${\bf{A}}_2$ is a Gaussian matrix, and $q_1=q_2=q$.}
 \label{figure2}
\end{figure}

\begin{figure}[!t]
 \centering
 \includegraphics[scale = 0.50]{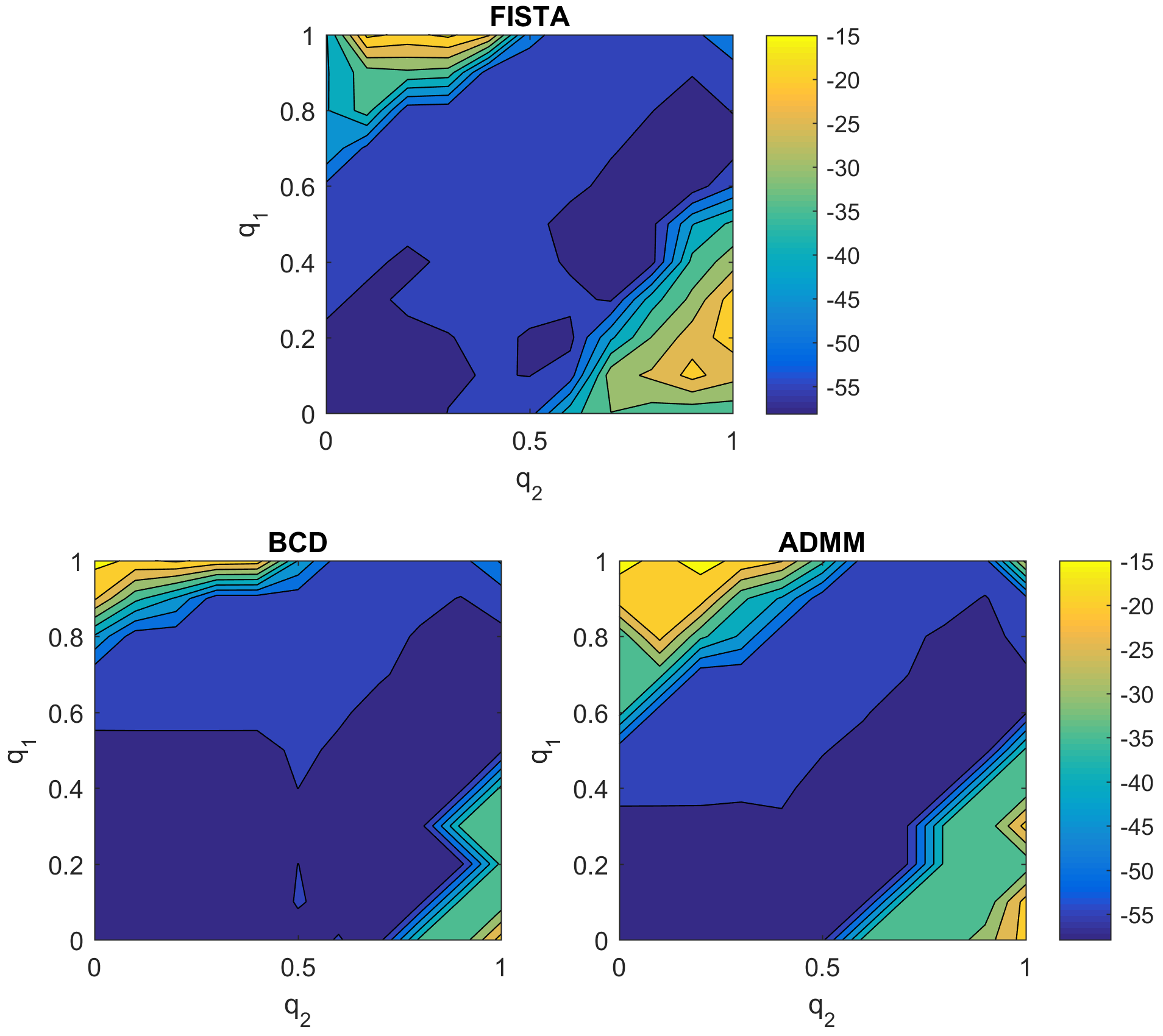}
\caption{Recovery performance of BCD and ADMM versus $q_1$ and $q_2$, in terms of RelErr in dB defined as $20{\log _{10}}(\|{\hat{{\bf{ x}}}_1} - {{\bf{x}}_1}\|_2/\|{{\bf{x}}_1}\|_2)$.}
 \label{figure3}
\end{figure}

\subsection{Color Image Inpainting}

\begin{figure*}[!t]
 \centering
 \includegraphics[scale = 0.30]{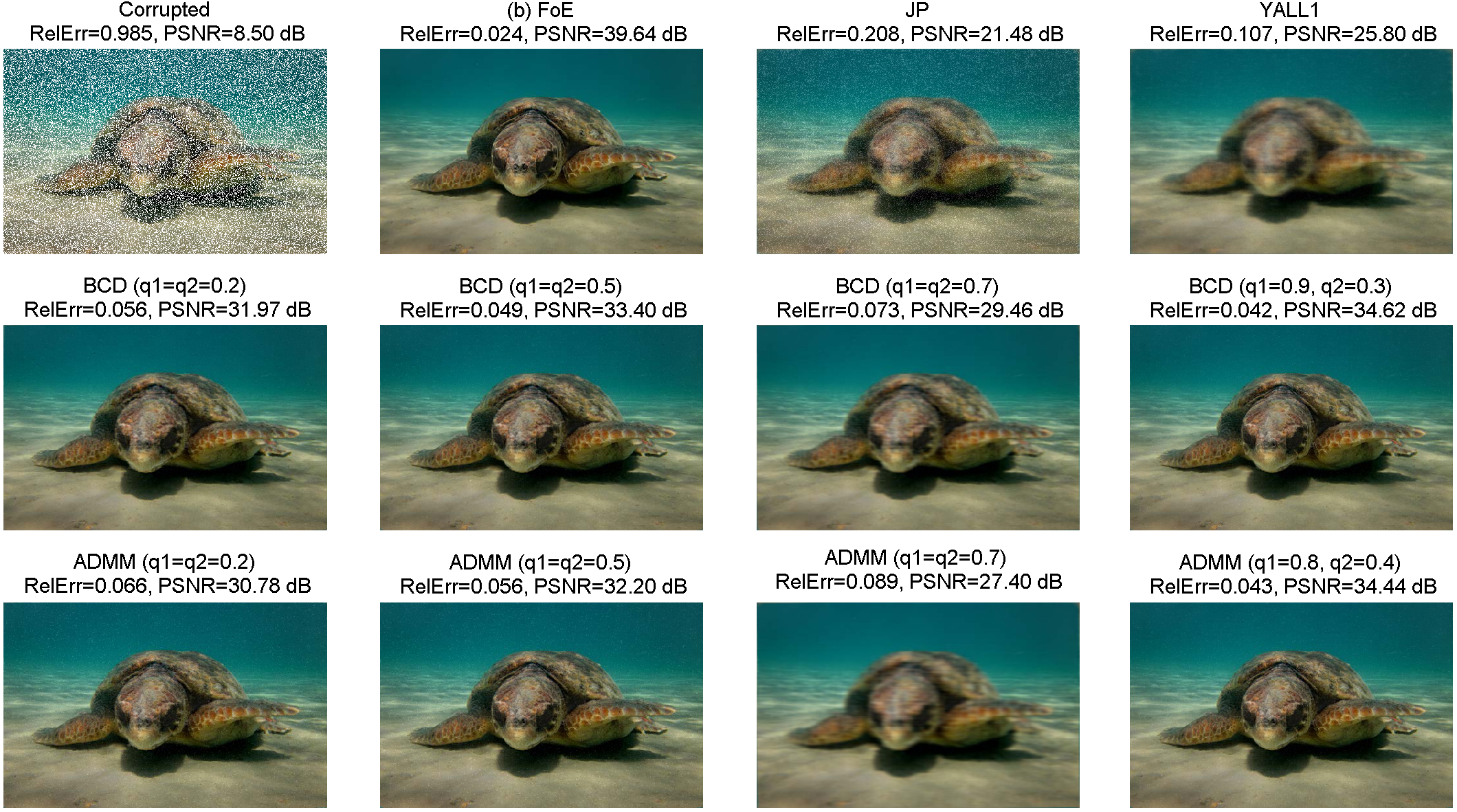}
\caption{Restoration of a $318 \times 500$ image corrupted by salt-and-pepper impulsive noise using the compared methods. (a) Corrupted image with salt-and-pepper impulsive noise (30\% of the pixels are corrupted). (b) FoE (PSNR = 39.64 dB). (c) JP (PSNR = 21.48 dB). (d) YALL1 (PSNR = 25.80 dB). (e)--(h) Proposed BCD method for different $q_1$ and $q_2$ (best PSNR = 34.62 dB). (i)--(l) Proposed ADMM method for different $q_1$ and $q_2$ (best PSNR = 34.44 dB). Note: Even though FoE is significantly effective, it requires the mask of the corruption. }
 \label{figure4}
\end{figure*}

In this subsection, we evaluate the performance of the new methods via inpainting experiments, in comparison with two existing $\ell_1$ solvers, JP [4] and YALL1 [34], and a classic inpainting method using the Field of Experts (FoE) model [50].
%YALL1 solves the robust $\ell_1$-LA formulation (5) using an ADMM scheme. JP is in fact a special case of YALL1 with $\mu=1$.
It has been shown in [49] that the k-SVD [49] and FoE [50] methods have comparable performance in color image inpainting, and both methods outperform the method in [9].
While there exist a number of inpainting methods in the literature, e.g., [7]--[9], [42], [43], the focus here is to quantify the impact of the values of $q_1$ and $q_2$ on the performance in comparison with the $\ell_1$-regularized methods.
The proposed BCD and ADMM algorithms are initialized by S-ADMM with $q_{1} = {q_2} = 1$ and $\mu=1$.

The goal is to separate the original image from sparse corruption. It is typically a sparse demixing problem of the form (1) with ${{\bf{A}}_1}$ be a basis of the image and ${{\bf{A}}_2} = {\bf{I}}$. We select ${{\bf{A}}_1}$ as an inverse discrete cosine transformation (IDCT) matrix, accordingly, ${{\bf{x}}_{\rm{1}}}$ is the DCT coefficients of the image. The advantage of using such a matrix is that the multiplication of ${{\bf{A}}_1}$ (or ${{\bf{A}}_1^T}$) with a vector can be rapidly obtained via IDCT (or DCT) of the vector, and thus scales well for high dimensional problems. The performance of the algorithms are evaluated in terms of RelErr of the estimated DCT coefficients ${{\bf{\hat x}}_1}$ and peak-signal noise ratio (PSNR) of the restored image.

We first consider an inpainting example in the presence of salt-and-pepper impulsive noise. 30\% of the pixels of the color image are corrupted by salt-and-pepper noise.
The multitask BCD and ADMM algorithms given in section IV are used to jointly recover the 3 channels of the color image. JP and YALL1 are also extended in a similar manner to the multitask case and used to jointly recover the 3 channels of the image. Fig. 4 shows the recovered images of the compared methods along with the RelErr and PSNR of each recovered image. Fig. 5 presents the recovery PSNR of the two proposed methods for different values of ${q_1}$ and ${q_2}$.

\begin{figure}[!t]
\centering
\includegraphics[scale = 0.53]{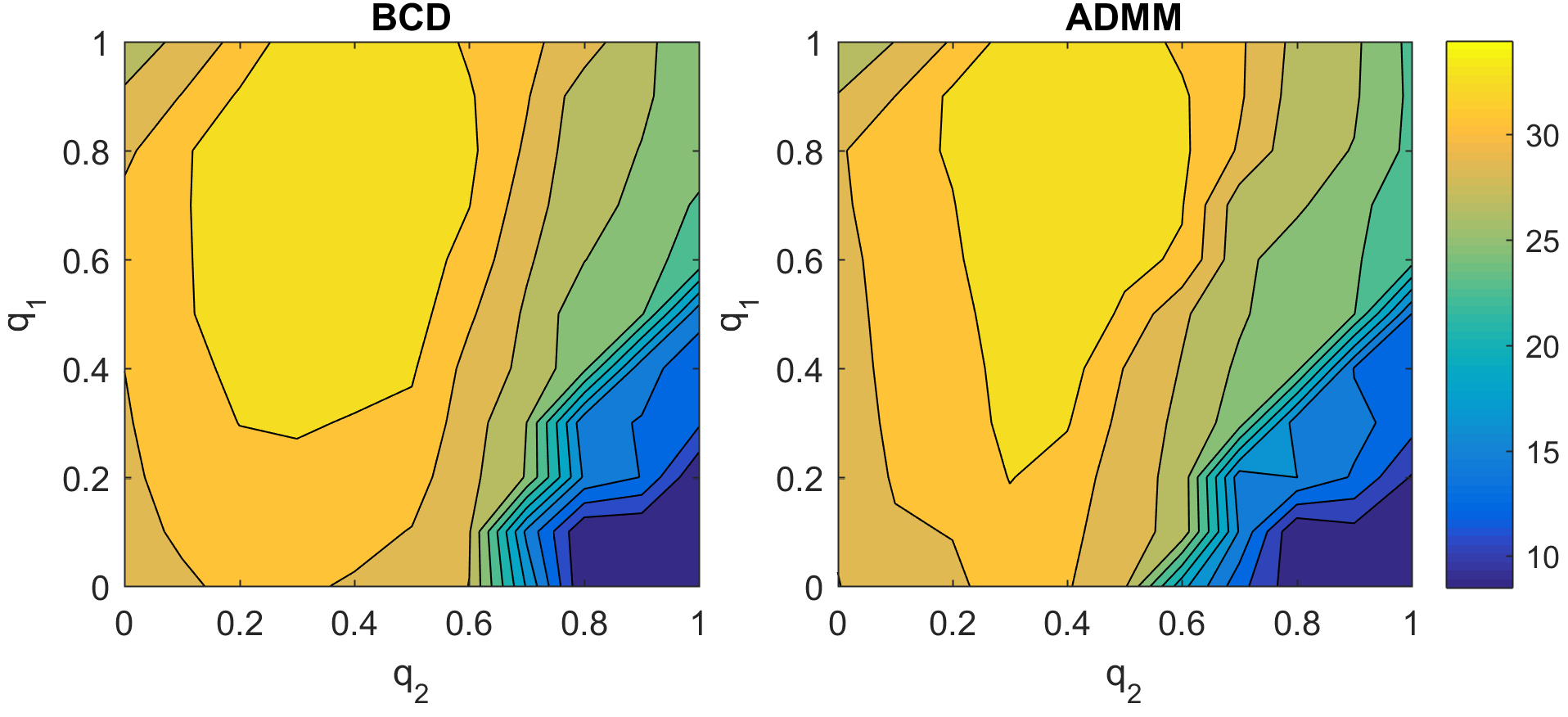}
\caption{Recovery performance of multitask BCD and ADMM versus $q_1$ and $q_2$ in color image inpainting corrupted by salt-and-pepper noise (in terms of PSNR in dB). The best BCD reconstruction (PSNR = 34.62 dB) is given by $q_1=0.9$ and $q_2=0.3$. The best ADMM reconstruction (PSNR = 34.44 dB) is given by $q_1=0.8$ and $q_2=0.4$.}
\label{figure5}
\end{figure}

\begin{figure*}[!t]
 \centering
 \includegraphics[scale = 0.47]{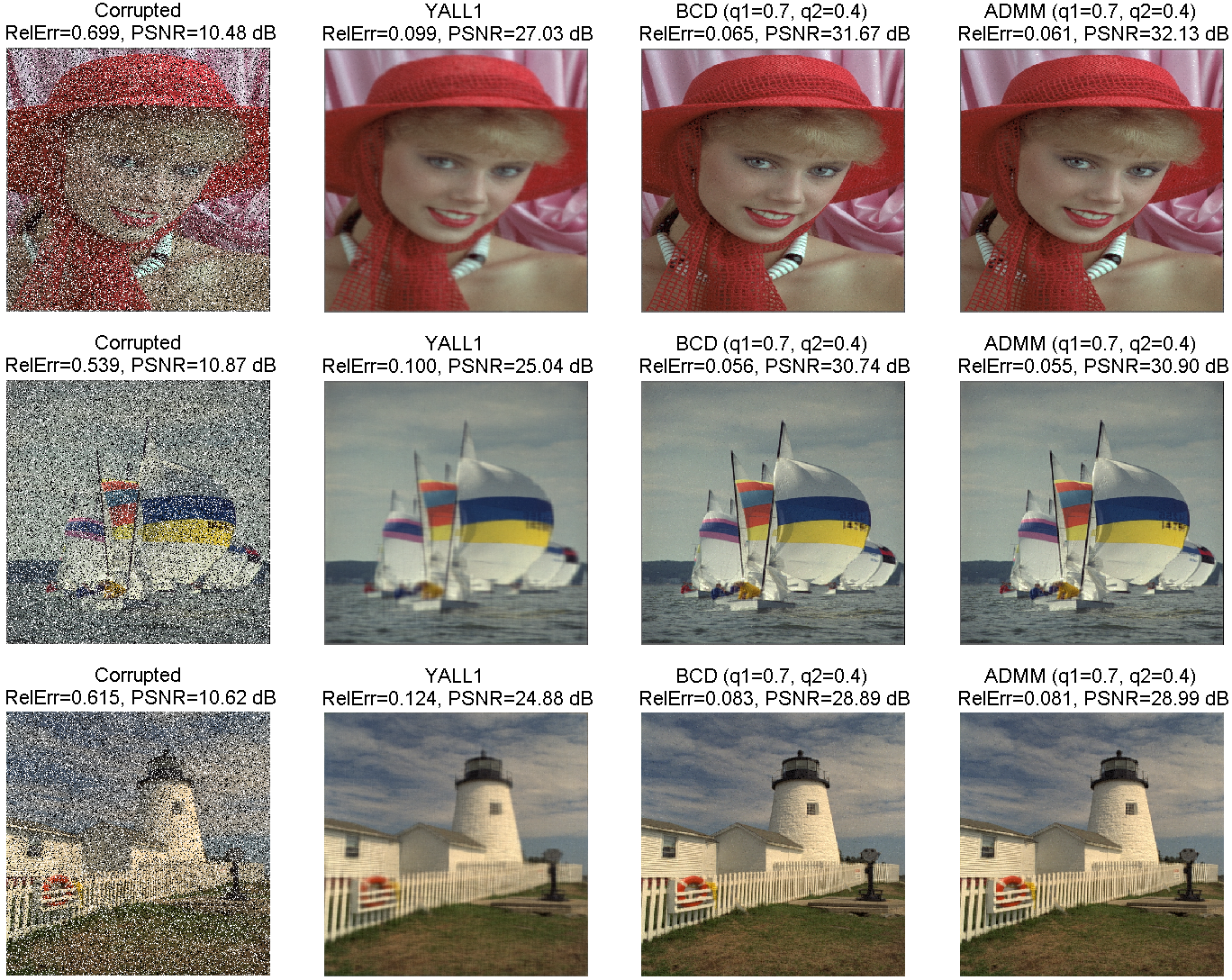}
\caption{Restoration of three $512 \times 512$ images corrupted by salt-and-pepper noise using the compared methods (30\% of the pixels are corrupted). The proposed BCD and ADMM methods at $q_1 = 0.7$ and $q_2=0.4$ outperform YALL1 reconstruction with 4-5 dB improvement.}
 \label{figure6}
\end{figure*}

It can be seen from Fig. 4 that, JP is outperformed by YALL1 and the new BCD and ADMM methods. It is reasonable since JP is a special case of YALL1 with $\mu  = 1$. Since YALL1 often attains its best performance at a value $\mu  \ne 1$, it outperforms JP in most cases. With properly chosen ${q_1}$ and ${q_2}$, both the new methods achieve surprisingly better recovery performance compared with the JP and YALL1 methods. FoE achieves the best performance and significantly outperforms our algorithms. However, while FoE (as well as the methods [9] and [49]) requires the exact support-set knowledge (mask) of the corruption, our algorithms do not use such prior information.

From Fig. 5, the best performance of BCD is given by ${q_{\rm{1}}} = 0.9$ and ${q_2} = 0.3$, which yields a recovery PSNR 8.8 dB higher than that of YALL1 (34.62 dB vs. 25.80 dB), with the corresponding RelErr be only approximately 39.3\% that of YALL1 (0.042 vs. 0.107). The best performance of ADMM is given by ${q_{\rm{1}}} = 0.8$ and ${q_2} = 0.4$, which yields a recovery PSNR 8.64 dB higher than that of YALL1 (34.44 dB vs. 25.80 dB), with the corresponding RelErr be only approximately 40.2\% that of YALL1 (0.043 vs. 0.107).
Moreover, the worst performance of both the BCD and ADMM algorithms are given by ${q_{\rm{1}}} = 0$ and ${q_2} = 1$. The results imply that, to attain a good inpainting performance, a moderate to large value should be used for ${q_{\rm{1}}}$, while a relatively small value should be used for ${q_{\rm{2}}}$. This is due the nature that, the DCT (also wavelet) coefficients ${{\bf{x}}_{\rm{1}}}$ of a real-life image are not strictly sparse but rather compressible, e.g., with DCT (also wavelet) coefficients approximately follow an exponential decay. But the considered corruption coefficients ${{\bf{x}}_2}$ are strictly sparse.

Table I compares the recovery results given by each algorithm in two conditions, the single-task condition and multitask condition. Unlike in the multitask condition each algorithm recovers the 3 channels of the image jointly, in the single-task condition each algorithm recovers the 3 channels independently. In the single-task condition, the BCD and ADMM algorithms given in section III are used. For the new algorithms, different values of $q_1$ and $q_2$ have been considered. From Table I, the multitask algorithms outperforms their single-task counterparts. This advantage can be expected to increase as the number of channels increases in some applications involving joint recovery.

Fig. 6 shows the recovery results on more example images (three $512\times512$ color images) in the presence of salt-and-pepper impulsive noise (30\% of the pixels are corrupted). For the proposed (multi-task) BCD and ADMM algorithms, we use $q_1=0.7$ and $q_2=0.4$. The results also demonstrate the significant improvement of nonconvex regularization over the $\ell_1$-regularization. Generally, the proposed BCD and ADMM algorithms have comparable performance.

In recovering the corrupted image in Fig. 4 and on a desktop PC with an Intel Core i7-4790K CPU at 4.0 GHz with 16 GB RAM, the runtime of FoE, JP and YALL1 (for a fixed $\mu$) are approximately 84, 29 and 26 seconds, respectively, while that of the proposed BCD and ADMM algorithms (for a fixed $\mu$) for different $q_1$ and $q_2$ ranges from 28 to 47 seconds.

\begin{table}[!t]
\caption{Recovery performance of the compared methods (Single-task: the 3 channels are independently recovered; Multitask: the 3 channels are jointly recovered).}
\footnotesize
\centering
\newcommand{\tabincell}[2]{\begin{tabular}{@{}#1@{}}#2\end{tabular}}
\begin{tabular}{|c|c|c|c|c|}
\hline
\multirow{2}{*}{Method} & \multicolumn{2}{|c|}{Single-task} & \multicolumn{2}{|c|}{Multitask}\\
\cline{2-5}
  &  RelErr & PSNR (dB) & RelErr & PSNR (dB)\\
\hline
JP & 0.256 & 20.21 & \textbf{0.208} & \textbf{21.48}\\
\hline
YALL1 & 0.110 &25.45 & \textbf{0.107} & \textbf{25.80}\\
\hline
\tabincell{c}{BCD\\($q_1=q_2=0.2$)} & 0.058 & 31.88 & \textbf{0.056} & \textbf{31.97}\\
\hline
\tabincell{c}{BCD\\($q_1=q_2=0.5$)} & 0.061 & 31.18 & \textbf{0.049} & \textbf{33.40}\\
\hline
\tabincell{c}{BCD\\ ($q_1=0.7, q_2=0.4$)} & 0.050 & 33.23 & \textbf{0.044} & \textbf{34.23}\\
\hline
\tabincell{c}{ADMM\\ ($q_1=q_2=0.2$)} & 0.068 & 30.24 & \textbf{0.066} & \textbf{30.78}\\
\hline
\tabincell{c}{ADMM\\ ($q_1=q_2=0.5$)} & 0.091 & 27.46 & \textbf{0.056} & \textbf{32.20}\\
\hline
\tabincell{c}{ADMM\\ ($q_1=0.7, q_2=0.4$)} & 0.048 & 33.53 & \textbf{0.043} & \textbf{34.39}\\
\hline
\end{tabular}
\end{table}

\subsection{Robust Compressive Sensing in Impulsive Noise}

In the last experiment, we consider the robust sparse recovery problem in compressive sensing in the presence of impulsive measurement noise. We use a simulated $K$-sparse signal ${\bf{x}}_1$ of length ${n_1} = 256$. The positions of the $K$ nonzeros are uniformly randomly chosen while the amplitude of each nonzero entry follows a Gaussian distribution. The $100 \times 256$ sensing matrix ${{\bf{A}}_1}$ is an orthonormal Gaussian random matrix. ${{\bf{A}}_2}$ is an identity matrix and ${\bf{x}}_2$ is symmetric $\alpha$-stable ($S\alpha S$) noise. Except for a few known cases, the $S\alpha S$ distributions do not have analytical formulations, but can be conveniently described by the characteristic function
\begin{equation}
\varphi (\omega ) = \exp \left( { - {\gamma ^\alpha }|\omega {|^\alpha }} \right)\notag
\end{equation}
where $0 < \alpha \le 2$ is the characteristic exponent and $\gamma  > 0$ is the scale parameter. The characteristic exponent measures the thickness of the tail of the distribution. The smaller the value of $\alpha$, the heavier the tail of the distribution and the more impulsive the noise is.

Fig. 7 shows the recovery performance of the proposed BCD and ADMM algorithms compared with YALL1. ${\bf{x}}_2$ is $S\alpha S$ noise with $\alpha = 1$ and $\gamma = {10^{ - 3}}$. The result is an average over 300 independent runs. It can be seen that both the proposed algorithms significantly outperforms YALL1. Fig. 8 presents the recovery performance of the proposed algorithms for different values of $q_1$ and $q_2$. The result indicates that a relatively small value of $q_1$ and a moderate to large value of $q_2$ should be used, e.g., ${q_{1}} \le 0.5$ and ${q_2} \ge 0.5$. This is reasonable because ${\bf{x}}_1$ is strictly sparse while the $S\alpha S$ noise ${\bf{x}}_2$ is not strictly sparse.

\begin{figure}[!t]
 \centering
 \includegraphics[scale = 0.20]{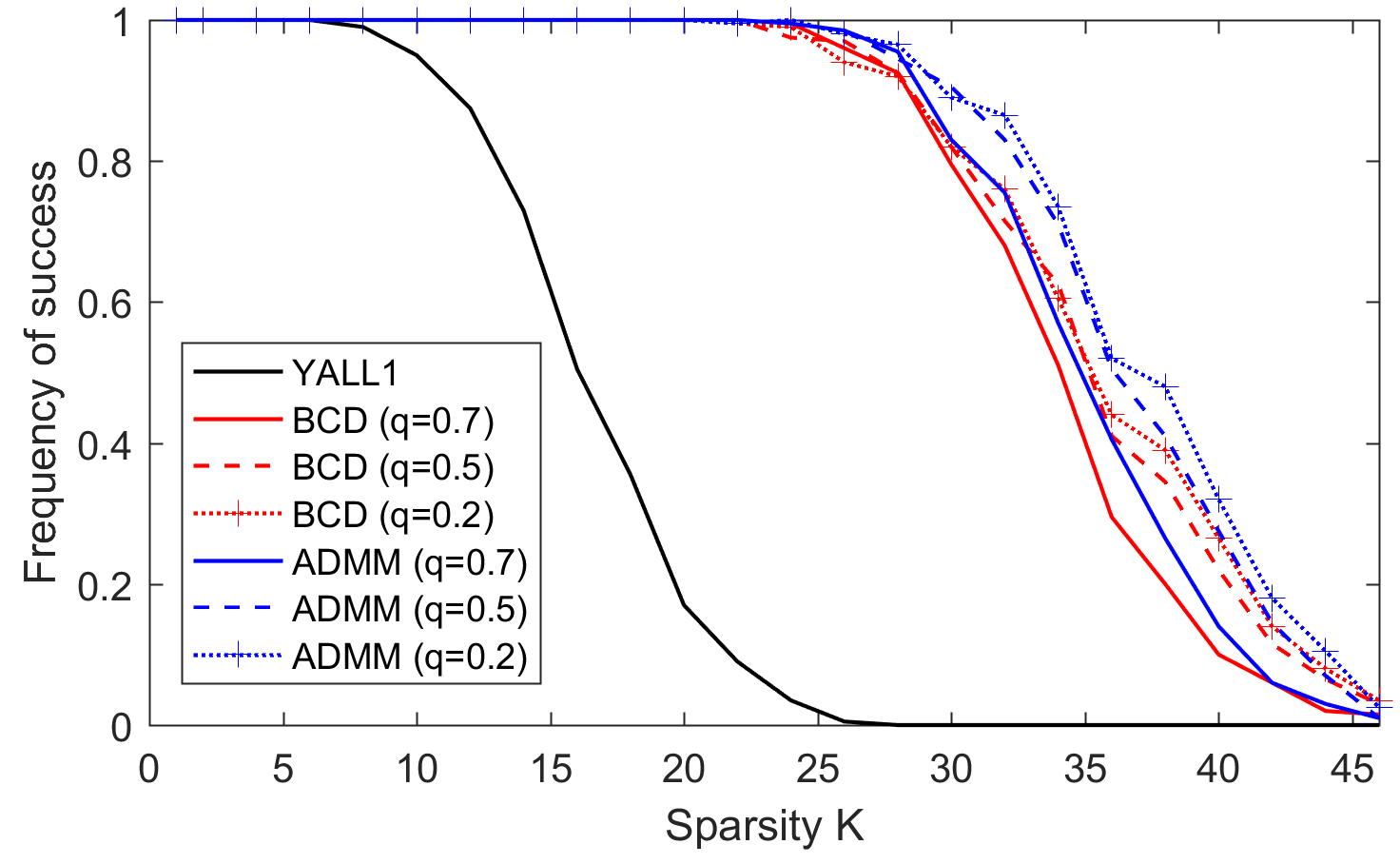}
\caption{Frequency of successful recovery versus sparsity, ${\bf{A}}_1$ is a Gaussian matrix, ${\bf{A}}_2$ is an identity matrix, ${{\bf{x}}_2}$ is $S\alpha S$ noise with $\alpha = 1$ and $\gamma = {10^{ - 3}}$, and $q_1=q_2=q$.}
 \label{figure7}
\end{figure}

\begin{figure}[!t]
 \centering
 \includegraphics[scale = 0.53]{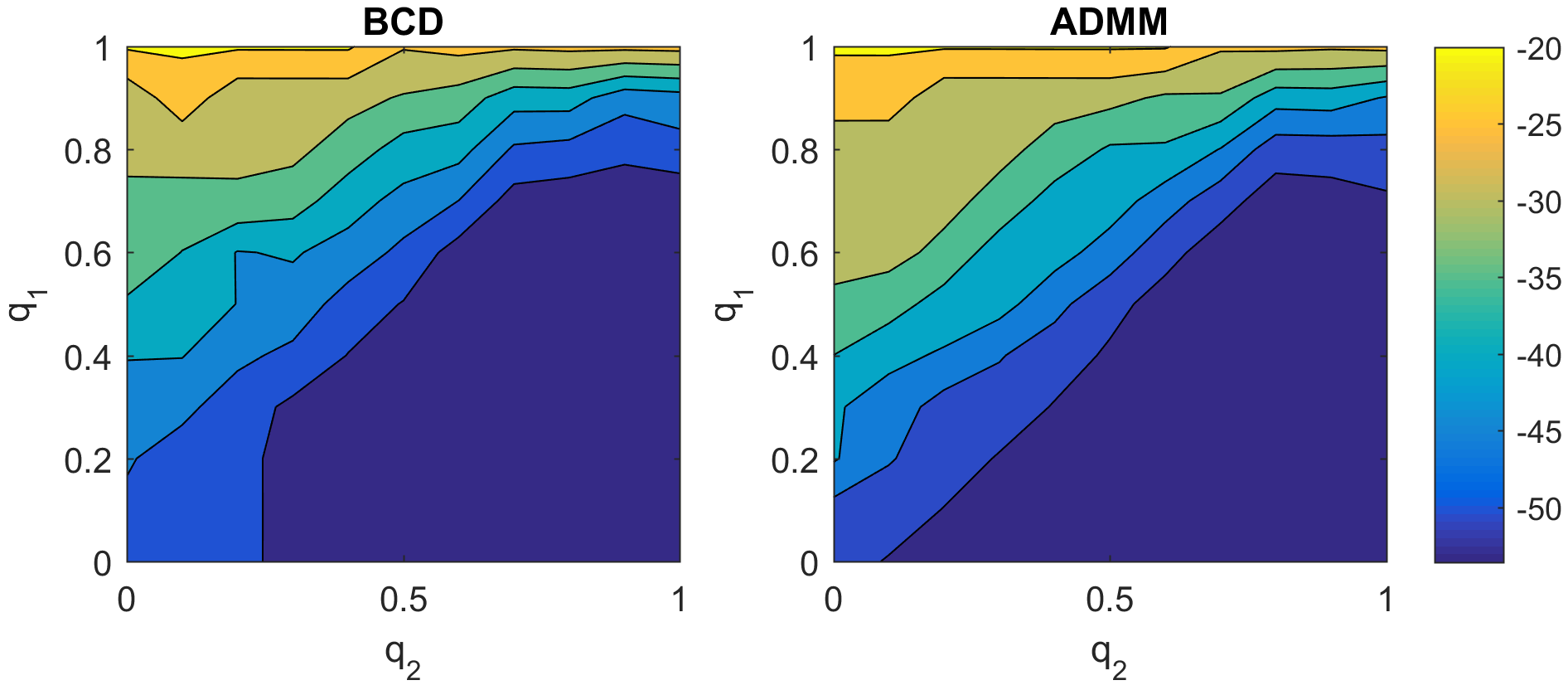}
\caption{Recovery performance of BCD and ADMM versus $q_1$ and $q_2$ in $S\alpha S$ noise with $\alpha = 1$ and $\gamma = {10^{ - 3}}$, in terms of RelErr in dB defined as $20{\log _{10}}(\|{\hat{{\bf{ x}}}_1} - {{\bf{x}}_1}\|_2/\|{{\bf{x}}_1}\|_2)$.}
 \label{figure8}
\end{figure}

\section{Conclusions}
In this paper, we proposed a novel formulation for sparse signals recovery and demixing using $\ell_q$-norm ($0 \le q < 1$) for sparsity inducing.
Two first-order algorithms have been developed to solve an approximation of this nonconvex $\ell_q$-minimization formulation.
The two algorithms are based on the BCD and ADMM frameworks, respectively, which are convergent under some mild conditions and scale well for high-dimensional problems.
Furthermore, the new algorithms have been extended for the multitask case.
Experiments demonstrated that the new algorithms can achieve considerable performance gain over the $\ell_1$-minimization algorithms. Moreover, by exploiting the multi-channel joint sparse pattern, the multitask versions of these methods can attain further performance improvement.

In practical applications, $q_1$ and  $q_2$ can be selected in an application-dependent manner.
For example, when $\ell_q$-norm is used as regularization for the DCT or wavelets coefficients of real-life images,
a moderate to large value of $q$ (e.g., $q \in [0.5,0.8]$) can yield good performance [52], which accords well with our results in inpainting experiments. This is due to the fact that the DCT (also wavelets) coefficients of a real-life image are not strictly sparse but rather follow an exponential decay. On the other hand, for strictly sparse signals, a relatively small value of $q$ would give good performance, e.g., $q \le 0.5$.

\appendices

\section{Proof of Theorem 2}
We first give the following lemmas in the proof of Theorem 2. In the sequel for convenience we use the notations: ${{\bf{v}}^k}: = ({\bf{x}}_{\rm{1}}^k,{\bf{x}}_2^k,{\bf{z}}_{\rm{1}}^k,{\bf{z}}_2^k,{\bf{w}}_{\rm{1}}^k,{\bf{w}}_2^k)$, ${\lambda _i} = {\lambda _{\max }}({\bf{A}}_i^T{{\bf{A}}_i})$ and ${\varphi _i} = {\lambda _{\min }}({\bf{A}}_i^T{{\bf{A}}_i})$, $i = 1,2$.

\textbf{Lemma 1.} Define $\tilde {\mathcal{L}}({{\bf{v}}^k},\tilde{\bf{ x}}): = \mathcal{L}({{\bf{v}}^k}) + {c_1}\left\| {{\bf{x}}_2^k - \tilde{\bf{ x}}} \right\|_{\rm{2}}^{\rm{2}}$ with ${c_1} = 8{\lambda _1}{\lambda _2}/{\rho _1}$, for the sequence $\{ {{\bf{v}}^k}\}$ generated via (18)-(23), if (26) holds, then
\begin{equation}
\begin{split}
&\tilde {\mathcal{L}}({{\bf{v}}^{k + 1}},{\bf{x}}_2^k) + {c_2}\left\| {{\bf{x}}_1^{k + 1} - {\bf{x}}_1^k} \right\|_2^2 + {c_3}\left\| {{\bf{x}}_2^{k + 1} - {\bf{x}}_2^k} \right\|_2^2\\
 &\le \tilde {\mathcal{L}}({{\bf{v}}^k},{\bf{x}}_2^{k - 1})\notag
\end{split}
\end{equation}
where ${c_2},{c_3} > 0$ are given by
\begin{equation}
\begin{split}
{c_2} &= \frac{{2{\varphi _1} + {\rho _1}}}{2} - \frac{{8\lambda _1^2}}{{{\rho _1}}} - \frac{{8{\lambda _1}{\lambda _2}}}{{{\rho _2}}},\\
{c_3} &= \frac{{2{\varphi _2} + {\rho _2}}}{2} - \frac{{8\lambda _2^2}}{{{\rho _2}}} - \frac{{8{\lambda _1}{\lambda _2}}}{{{\rho _1}}}. \notag
\end{split}
\end{equation}

\textbf{Lemma 2.} For the sequence $\{ {{\bf{v}}^k}\}$ generated via (18)--(23), if (26) holds, then
\begin{equation}
\mathop {\lim }\limits_{k \to \infty } \left\| {{{\bf{v}}^{k + 1}} - {{\bf{v}}^k}} \right\|_2^2 = 0.\notag
\end{equation}
In particular, any cluster point of $\{ {{\bf{v}}^k}\}$ is a stationary point of ${\mathcal{L}}$.

\textbf{Lemma 3.} For $\tilde {\mathcal{L}}({{\bf{v}}^k},\tilde{\bf{ x}}): = {\mathcal{L}}({{\bf{v}}^k}) + {c_1}\left\| {{\bf{x}}_2^k - \tilde{\bf{ x}}} \right\|_{\rm{2}}^{\rm{2}}$ as defined in Lemma 1, for the sequence $\{ {{\bf{v}}^k}\}$ generated via (18)--(23), suppose that (26) holds, then there exists a constant ${c_4} > 0$ such that
\begin{equation}
\begin{split}
&{\rm{dist}}(0,\partial \tilde L({{\bf{v}}^{k{\rm{ + }}1}},{\bf{x}}_2^k))\\
& \le {c_4}({\left\| {{\bf{x}}_1^{k + 1} - {\bf{x}}_1^k} \right\|_2} + {\left\| {{\bf{x}}_2^{k + 1} - {\bf{x}}_2^k} \right\|_2} + {\left\| {{\bf{x}}_2^k - {\bf{x}}_2^{k - 1}} \right\|_2}).\notag
\end{split}
\end{equation}

\textbf{Proof of Lemma 1:} First, the minimizer ${\bf{x}}_1^{k+1}$ given by (20) satisfies
\begin{equation} %(40)
2{\bf{A}}_1^T({{\bf{A}}_1}{\bf{x}}_1^{k + 1} + {{\bf{A}}_2}{\bf{x}}_2^k - {\bf{y}}) + {\rho _1}({\bf{x}}_1^{k + 1} - {\bf{z}}_1^{k + 1} + {\bf{w}}_1^k/{\rho _1}) = {\bf{0}}.
\end{equation}
Substituting (22) into (40) yields
\begin{equation} %(41)
{\bf{w}}_1^{k + 1} =  - 2{\bf{A}}_1^T({{\bf{A}}_1}{\bf{x}}_1^{k + 1} + {{\bf{A}}_2}{\bf{x}}_2^k - {\bf{y}}).
\end{equation}
Then, it follows from (41) that
\begin{equation} %(42)
\begin{split}
&\left\| {{\bf{w}}_1^{k + 1} - {\bf{w}}_1^k} \right\|_2^2\\
& = 4\left\| {{\bf{A}}_1^T{{\bf{A}}_1}({\bf{x}}_1^{k + 1} - {\bf{x}}_1^k) + {\bf{A}}_1^T{{\bf{A}}_2}({\bf{x}}_2^k - {\bf{x}}_2^{k - 1})} \right\|_2^2\\
& \le 4{\big( {{{\left\| {{\bf{A}}_1^T{{\bf{A}}_1}({\bf{x}}_1^{k + 1} - {\bf{x}}_1^k)} \right\|}_2} + {{\left\| {{\bf{A}}_1^T{{\bf{A}}_2}({\bf{x}}_2^k - {\bf{x}}_2^{k - 1})} \right\|}_2}} \big)^2}\\
& \le 8\lambda _1^2\left\| {{\bf{x}}_1^{k + 1} - {\bf{x}}_1^k} \right\|_2^2 + 8{\lambda _1}{\lambda _2}\left\| {{\bf{x}}_2^k - {\bf{x}}_2^{k - 1}} \right\|_2^2
\end{split}
\end{equation}
where $\lambda _{\max }^2({\bf{A}}_1^T{{\bf{A}}_2}) \le {\lambda _1}{\lambda _2}$ is used for the last inequality. Similarly, from the definition of ${\bf{x}}_2^{k+1}$ as a minimizer of (21), and with the use of (23), we have
\begin{equation} %(43)
{\bf{w}}_2^{k + 1} =  - 2{\bf{A}}_2^T({{\bf{A}}_1}{\bf{x}}_1^{k + 1} + {{\bf{A}}_2}{\bf{x}}_2^{k + 1} - {\bf{y}})
\end{equation}
and further
\begin{equation} %(44)
\begin{split}
&\left\| {{\bf{w}}_2^{k + 1} - {\bf{w}}_2^k} \right\|_2^2\\
& \le 8{\lambda _1}{\lambda _2}\left\| {{\bf{x}}_1^{k + 1} - {\bf{x}}_1^k} \right\|_2^2 + 8\lambda _2^2\left\| {{\bf{x}}_2^{k + 1} - {\bf{x}}_2^k} \right\|_2^2.
\end{split}
\end{equation}

From (22), (23) and the definition of ${\mathcal{L}}$, we have
\begin{equation} %(45)
\begin{split}
&{\mathcal{L}}({\bf{x}}_{\rm{1}}^{k + 1},{\bf{x}}_2^{k + 1},{\bf{z}}_{\rm{1}}^{k + 1},{\bf{z}}_2^{k + 1},{\bf{w}}_{\rm{1}}^{k + 1},{\bf{w}}_2^k)\\
& - {\mathcal{L}}({\bf{x}}_{\rm{1}}^{k + 1},{\bf{x}}_2^{k + 1},{\bf{z}}_{\rm{1}}^{k + 1},{\bf{z}}_2^{k + 1},{\bf{w}}_{\rm{1}}^k,{\bf{w}}_2^k) = \frac{1}{{{\rho _1}}}\left\| {{\bf{w}}_1^{k + 1} - {\bf{w}}_1^k} \right\|_2^2
\end{split}
\end{equation}
and
\begin{equation} %(46)
\begin{split}
&{\mathcal{L}}({\bf{x}}_{\rm{1}}^{k + 1},{\bf{x}}_2^{k + 1},{\bf{z}}_{\rm{1}}^{k + 1},{\bf{z}}_2^{k + 1},{\bf{w}}_{\rm{1}}^{k + 1},{\bf{w}}_2^{k + 1})\\
& \!-\! {\mathcal{L}}({\bf{x}}_{\rm{1}}^{k + 1},{\bf{x}}_2^{k + 1},{\bf{z}}_{\rm{1}}^{k + 1},{\bf{z}}_2^{k + 1},{\bf{w}}_{\rm{1}}^{k + 1},{\bf{w}}_2^k) \!=\! \frac{1}{{{\rho _2}}}\left\| {{\bf{w}}_2^{k + 1} \!-\! {\bf{w}}_2^k} \right\|_2^2.
\end{split}
\end{equation}
Since $\mathcal{L}({{\bf{x}}_1},{\bf{x}}_2^k,{\bf{z}}_{\rm{1}}^{k + 1},{\bf{z}}_2^{k + 1},{\bf{w}}_{\rm{1}}^k,{\bf{w}}_2^k)$ is $(2{\varphi _1} + {\rho _1})$-strongly convex, for any ${\bf{x}}_{\rm{1}}^k \in {\mathbb{R}^{{n_1}}}$, the minimizer ${\bf{x}}_1^{k{\rm{ + }}1}$ given by (20) satisfies
\begin{equation} %(47)
\begin{split}
&{\mathcal{L}}({\bf{x}}_{\rm{1}}^{k + 1},{\bf{x}}_2^k,{\bf{z}}_{\rm{1}}^{k + 1},{\bf{z}}_2^{k + 1},{\bf{w}}_{\rm{1}}^k,{\bf{w}}_2^k)\\
& \le {\mathcal{L}}({\bf{x}}_{\rm{1}}^k,{\bf{x}}_2^k,{\bf{z}}_{\rm{1}}^{k + 1},{\bf{z}}_2^{k + 1},{\bf{w}}_{\rm{1}}^k,{\bf{w}}_2^k)\!-\! \frac{{2{\varphi _1}\! +\! {\rho _1}}}{2}\left\| {{\bf{x}}_{\rm{1}}^{k + 1} - {\bf{x}}_{\rm{1}}^k} \right\|_2^2.
\end{split}
\end{equation}
Similarly, as ${\mathcal{L}}({\bf{x}}_{\rm{1}}^{k + 1},{{\bf{x}}_2},{\bf{z}}_{\rm{1}}^{k + 1},{\bf{z}}_2^{k + 1},{\bf{w}}_{\rm{1}}^k,{\bf{w}}_2^k)$ is $(2{\varphi _2} + {\rho _2})$-strongly convex, for any ${\bf{x}}_2^k \in {\mathbb{R}^{{n_2}}}$, the minimizer ${\bf{x}}_2^{k{\rm{ + }}1}$ given by (21) satisfies
\begin{equation} %(48)
\begin{split}
&{\mathcal{L}}({\bf{x}}_{\rm{1}}^{k + 1},{\bf{x}}_2^{k + 1},{\bf{z}}_{\rm{1}}^{k + 1},{\bf{z}}_2^{k + 1},{\bf{w}}_{\rm{1}}^k,{\bf{w}}_2^k)\\
& \le {\mathcal{L}}({\bf{x}}_{\rm{1}}^{k + 1},{\bf{x}}_2^k,{\bf{z}}_{\rm{1}}^{k + 1},{\bf{z}}_2^{k + 1},{\bf{w}}_{\rm{1}}^k,{\bf{w}}_2^k)\\
&~~~~~~~~- \frac{{2{\varphi _2} + {\rho _2}}}{2}\left\| {{\bf{x}}_2^{k + 1} - {\bf{x}}_2^k} \right\|_2^2.
\end{split}
\end{equation}
Moreover, the minimizer ${\bf{z}}_{\rm{1}}^{k + 1}$ given by (18) satisfies
\begin{equation} %(49)
\begin{split}
&{\mathcal{L}}({\bf{x}}_{\rm{1}}^k,{\bf{x}}_2^k,{\bf{z}}_{\rm{1}}^{k + 1},{\bf{z}}_2^k,{\bf{w}}_{\rm{1}}^k,{\bf{w}}_2^k) \le {\mathcal{L}}({\bf{x}}_{\rm{1}}^k,{\bf{x}}_2^k,{\bf{z}}_{\rm{1}}^k,{\bf{z}}_2^k,{\bf{w}}_{\rm{1}}^k,{\bf{w}}_2^k).
\end{split}
\end{equation}
Meanwhile, the minimizer ${\bf{z}}_2^{k + 1}$ given by (19) satisfies
\begin{equation} %(50)
\begin{split}
&{\mathcal{L}}({\bf{x}}_{\rm{1}}^k,{\bf{x}}_2^k,{\bf{z}}_{\rm{1}}^{k + 1},{\bf{z}}_2^{k + 1},{\bf{w}}_{\rm{1}}^k,{\bf{w}}_2^k)\!\le\! {\mathcal{L}}({\bf{x}}_{\rm{1}}^k,{\bf{x}}_2^k,{\bf{z}}_{\rm{1}}^{k + 1},{\bf{z}}_2^k,{\bf{w}}_{\rm{1}}^k,{\bf{w}}_2^k).
\end{split}
\end{equation}

Then, summing (45)--(50) and using (42) and (44) yields
\begin{equation} %(51)
\begin{split}
&{\mathcal{L}}({{\bf{v}}^{k + 1}}) - L({{\bf{v}}^k}) \\
&\le \left( {\frac{{8\lambda _1^2}}{{{\rho _1}}} + \frac{{8{\lambda _1}{\lambda _2}}}{{{\rho _2}}} - \frac{{2{\varphi _1} + {\rho _1}}}{2}} \right)\left\| {{\bf{x}}_1^{k + 1} - {\bf{x}}_1^k} \right\|_2^2\\
&~~~~ + \left( {\frac{{8\lambda _2^2}}{{{\rho _2}}} - \frac{{2{\varphi _2} + {\rho _2}}}{2}} \right)\left\| {{\bf{x}}_2^{k + 1} - {\bf{x}}_2^k} \right\|_2^2\\
&~~~~+ \frac{{8{\lambda _1}{\lambda _2}}}{{{\rho _1}}}\left\| {{\bf{x}}_2^k - {\bf{x}}_2^{k - 1}} \right\|_2^2
\end{split}
\end{equation}
which consequently results in Lemma 1, where ${c_2} > 0$ and ${c_3} > 0$ when (26) holds. This result indicates the auxiliary function $\tilde {\mathcal{L}}({{\bf{v}}^k},{\bf{x}}_2^{k - 1})$ is decreasing when (26) is satisfied.

\textbf{Proof of Lemma 2:} First, we show that, under the condition of (26), the sequence $\{ {{\bf{v}}^k}\}$ generated via (18)--(23) is bounded. It follows from (41) that
\begin{equation} %(52)
\begin{split}
&\left\| {{\bf{w}}_1^k} \right\|_2^2\\
& = 4\left\| {{\bf{A}}_1^T({{\bf{A}}_1}{\bf{x}}_1^k + {{\bf{A}}_2}{\bf{x}}_2^k - {\bf{y}}) - {\bf{A}}_1^T{{\bf{A}}_2}({\bf{x}}_2^k - {\bf{x}}_2^{k - 1})} \right\|_2^2\\
& \le 4{\left( {{{\left\| {{\bf{A}}_1^T({{\bf{A}}_1}{\bf{x}}_1^k \!+\! {{\bf{A}}_2}{\bf{x}}_2^k \!-\! {\bf{y}})} \right\|}_2} + {{\left\| {{\bf{A}}_1^T{{\bf{A}}_2}({\bf{x}}_2^k \!-\! {\bf{x}}_2^{k - 1})} \right\|}_2}} \right)^2}\\
& \le \frac{16}{3}{\lambda _1}\left\| {{{\bf{A}}_1}{\bf{x}}_1^k + {{\bf{A}}_2}{\bf{x}}_2^k - {\bf{y}}} \right\|_2^2 + 16{\lambda _1}{\lambda _2}\left\| {{\bf{x}}_2^k - {\bf{x}}_2^{k - 1}} \right\|_2^2
\end{split}
\end{equation}
where ${\lambda _{\max }}({{\bf{A}}_1}{\bf{A}}_1^T) = {\lambda _{\max }}({\bf{A}}_1^T{{\bf{A}}_1}) = {\lambda _1}$ and $\lambda _{\max }^2({\bf{A}}_1^T{{\bf{A}}_2}) \le {\lambda _1}{\lambda _2}$ are used for the last equality. Meanwhile, it follows from (43) that
\begin{equation} %(53)
\left\| {{\bf{w}}_2^k} \right\|_2^2 \le 4{\lambda _2}\left\| {{{\bf{A}}_1}{\bf{x}}_1^k + {{\bf{A}}_2}{\bf{x}}_2^k - {\bf{y}}} \right\|_2^2.
\end{equation}
Define ${\tilde{\bf{ v}}^k}: = ({\bf{x}}_{\rm{1}}^k,{\bf{x}}_2^k,{\bf{z}}_{\rm{1}}^k,{\bf{z}}_2^k,{\bf{w}}_{\rm{1}}^k,{\bf{w}}_2^k,{\bf{x}}_2^{k - 1})$ and $\tilde {\mathcal{L}}({\tilde{\bf{ v}}^k}): = {\mathcal{L}}({{\bf{v}}^k}) + {c_1}\left\| {{\bf{x}}_2^k - {\bf{x}}_2^{k - 1}} \right\|_{\rm{2}}^{\rm{2}}$, since $\tilde {\mathcal{L}}({\tilde{\bf{ v}}^k})$ is lower semi-continuous, it is bounded from below. Meanwhile, from Lemma 1, when (26) is satisfied, $\tilde {\mathcal{L}}({\tilde{\bf{ v}}^k})$ is nonincreasing, thus it is convergent. From the definition of $\tilde {\mathcal{L}}$ and using (52) and (53), for any $k>1$ we have
\begin{equation} %(54)
\begin{split}
&\tilde {\mathcal{L}}({{\tilde{\bf{ v}}}^1}) \ge \tilde {\mathcal{L}}({{\tilde{\bf{ v}}}^k})\\
& = \left\| {{{\bf{A}}_1}{\bf{x}}_1^k + {{\bf{A}}_2}{\bf{x}}_2^k - {\bf{y}}} \right\|_2^2 + \frac{{8{\lambda _1}{\lambda _2}}}{{{\rho _1}}}\left\| {{\bf{x}}_2^k - {\bf{x}}_2^{k - 1}} \right\|_2^2\\
&+ \frac{{{\rho _1}}}{2}\left\| {{\bf{x}}_1^k - {\bf{z}}_1^k + \frac{{{\bf{w}}_1^k}}{{{\rho _1}}}} \right\|_2^2 + \frac{{{\rho _2}}}{2}\left\| {{\bf{x}}_2^k - {\bf{z}}_2^k + \frac{{{\bf{w}}_2^k}}{{{\rho _2}}}} \right\|_2^2\\
&- \frac{{\left\| {{\bf{w}}_1^k} \right\|_2^2}}{{2{\rho _1}}} - \frac{{\left\| {{\bf{w}}_2^k} \right\|_2^2}}{{2{\rho _2}}} + \beta \mu \left\| {{\bf{z}}_1^k} \right\|_{{q_1}}^{{q_1}} + \beta \left\| {{\bf{z}}_2^k} \right\|_{{q_2}}^{{q_2}}\\
& \ge {c_6}\left\| {{{\bf{A}}_1}{\bf{x}}_1^k + {{\bf{A}}_2}{\bf{x}}_2^k - {\bf{y}}} \right\|_2^2 + \beta \mu \left\| {{\bf{z}}_1^k} \right\|_{{q_1}}^{{q_1}} + \beta \left\| {{\bf{z}}_2^k} \right\|_{{q_2}}^{{q_2}}\\
& + \frac{{{\rho _1}}}{2}\left\| {{\bf{x}}_1^k - {\bf{z}}_1^k + \frac{{{\bf{w}}_1^k}}{{{\rho _1}}}} \right\|_2^2 + \frac{{{\rho _2}}}{2}\left\| {{\bf{x}}_2^k - {\bf{z}}_2^k + \frac{{{\bf{w}}_2^k}}{{{\rho _2}}}} \right\|_2^2
\end{split}
\end{equation}
where ${c_6} = 1 - \frac{8\lambda _1}{3\rho _1} -\frac{2\lambda _2}{\rho _2}$. When ${c_6} > 0$, from (52), (53) and (54), it is easy to see that the sequence $\{ {{\bf{v}}^k}\}$ is bounded. It can be proved after straightforward algebraic manipulation that ${c_6} > 0$ when (26) is satisfied. Briefly, let $x = \frac{\lambda _1}{\rho _1}$, $y = \frac{\lambda _2}{\rho _2}$, using ${\varphi _{\rm{1}}} \le {\lambda _{\rm{1}}}$ and ${\varphi _{\rm{2}}} \le {\lambda _{\rm{2}}}$, (26) can be rewritten as
\begin{align} %(55) (56)
0 < y < \frac{1}{{16x}} - x + \frac{1}{8},\\
0 < x < \frac{1}{{16y}} - y + \frac{1}{8}.
\end{align}
The domain of $(x,y)$ satisfies these two inequalities is shown as the green area in Fig. 9. In this domain the maximal value of $f(x,y) = {\frac{8}{3}}x{+}2y$ is given by $f(0.2108,0.2108) = {\rm{0.9837}}$. Thus, ${c_6} > 0$ when (26) is satisfied.

\begin{figure}[!t]
 \centering
 \includegraphics[scale = 0.23]{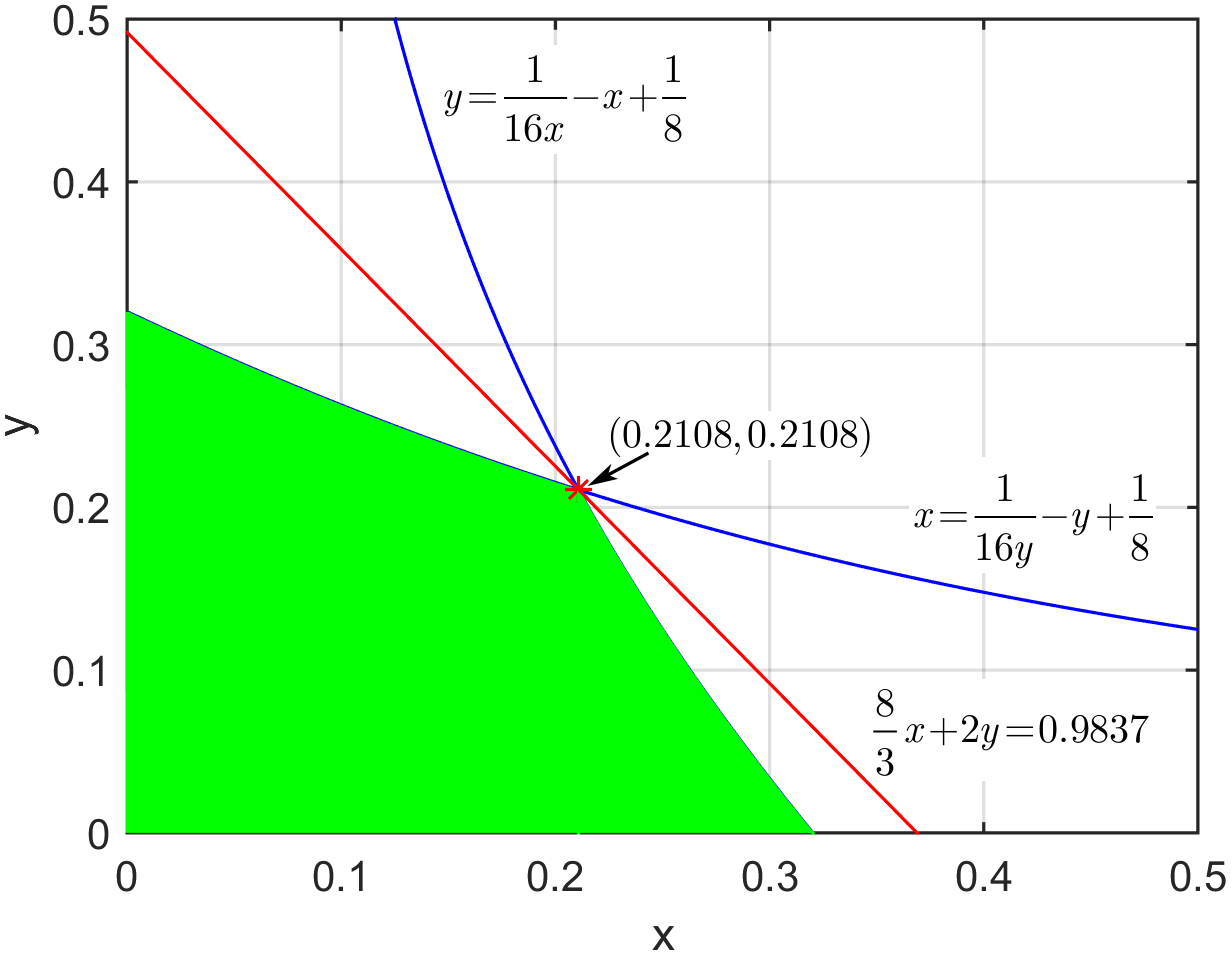}
\caption{Illustration of the maximal value of $f(x,y) = \frac{8}{3}x{+}2y$ under the two constraints (55) and (56).}
 \label{figure9}
\end{figure}

When ${\tilde{\bf{ v}}^k}$ is bounded, there exists a convergent subsequence ${\tilde{\bf{ v}}^{{k_j}}}$ which converges to a cluster point ${\tilde{\bf{ v}}^*}$. Further, when ${c_2} > 0$ and ${c_3} > 0$, it follows from Lemma 1 that $\tilde {\mathcal{L}}({\tilde{\bf{ v}}^k})$ is nonincreasing and convergent, and $\tilde {\mathcal{L}}({\tilde{\bf{ v}}^k}) \ge \tilde {\mathcal{L}}({\tilde{\bf{ v}}^*})$ for any $k\ge 1$. In this condition, from Lemma 1 we have
\begin{equation} %
\begin{split}
\infty  &> \tilde {\mathcal{L}}({{\tilde{\bf{ v}}}^1}) - \tilde L({{\tilde{\bf{ v}}}^*}) \ge \tilde {\mathcal{L}}({{\tilde{\bf{ v}}}^1}) - \tilde L({{\tilde{\bf{ v}}}^{N + 1}})\\
& = \sum\limits_{k = 1}^N {\left[ {\tilde L({{\tilde{\bf{ v}}}^k}) - \tilde L({{\tilde{\bf{ v}}}^{k + 1}})} \right]} \\
& \ge {c_2}\sum\limits_{k = 1}^N {\left\| {{\bf{x}}_1^{k + 1} - {\bf{x}}_1^k} \right\|_2^2}  + {c_3}\sum\limits_{k = 1}^N {\left\| {{\bf{x}}_2^{k + 1} - {\bf{x}}_2^k} \right\|_2^2}.\notag
\end{split}
\end{equation}
Let $N \to \infty $, when ${c_2} > 0$ and ${c_3} > 0$, we have
\begin{equation} %(57)
\sum\limits_{k = 1}^\infty  {\left\| {{\bf{x}}_1^{k + 1} - {\bf{x}}_1^k} \right\|_2^2}  < \infty ~~ {\rm{and}}~~
\sum\limits_{k = 1}^\infty  {\left\| {{\bf{x}}_2^{k + 1} - {\bf{x}}_2^k} \right\|_2^2}  < \infty
\end{equation}
which together with (42) and (44) implies
\begin{equation} %(58)
\sum\limits_{k = 1}^\infty  {\left\| {{\bf{w}}_1^{k + 1} - {\bf{w}}_1^k} \right\|_2^2}  < \infty ~~ {\rm{and}}~~
\sum\limits_{k = 1}^\infty  {\left\| {{\bf{w}}_2^{k + 1} - {\bf{w}}_2^k} \right\|_2^2}  < \infty.
\end{equation}
Moreover, based on (57), (58) and using (22), (23), we have
\begin{equation} %(59)
\sum\limits_{k = 1}^\infty  {\left\| {{\bf{z}}_1^{k + 1} - {\bf{z}}_1^k} \right\|_2^2}  < \infty ~~ {\rm{and}}~~
\sum\limits_{k = 1}^\infty  {\left\| {{\bf{z}}_2^{k + 1} - {\bf{z}}_2^k} \right\|_2^2}  < \infty.
\end{equation}
Then, from (57), (58) and (59), it is easy to see that $\mathop {\lim }\limits_{k \to \infty } \left\| {{{\bf{v}}^{k + 1}} - {{\bf{v}}^k}} \right\|_2^2 = 0$.

Next, we show that any cluster point of the sequence $\{ {{\bf{v}}^k}\}$ generated via (18)--(23) is a stationary point of $\mathcal{L}$. From the optimality conditions, the sequence generated via (18)--(23) satisfies
\begin{equation} %(60)
\left\{ {\begin{array}{*{20}{l}}
{{\bf{0}} \in \beta \mu \partial \left\| {{\bf{z}}_1^{k + 1}} \right\|_{{q_1}}^{{q_1}} - {\bf{w}}_1^{k + 1} + {\rho _1}({\bf{x}}_1^{k + 1} - {\bf{x}}_1^k)}\\
{{\bf{0}} \in \beta \partial \left\| {{\bf{z}}_2^{k + 1}} \right\|_{{q_2}}^{{q_2}} - {\bf{w}}_2^{k + 1} + {\rho _2}({\bf{x}}_2^{k + 1} - {\bf{x}}_2^k)}\\
{{\bf{0}} = {\bf{w}}_1^{k + 1} + 2{\bf{A}}_1^T({{\bf{A}}_1}{\bf{x}}_1^{k + 1} + {{\bf{A}}_2}{\bf{x}}_2^k - {\bf{y}})}\\
{{\bf{0}} = {\bf{w}}_2^{k + 1} + 2{\bf{A}}_2^T({{\bf{A}}_1}{\bf{x}}_1^{k + 1} + {{\bf{A}}_2}{\bf{x}}_2^{k + 1} - {\bf{y}})}\\
{{\bf{w}}_1^{k + 1} = {\bf{w}}_1^k + {\rho _1}({\bf{x}}_1^{k + 1} - {\bf{z}}_1^{k + 1})}\\
{{\bf{w}}_2^{k + 1} = {\bf{w}}_2^k + {\rho _2}({\bf{x}}_2^{k + 1} - {\bf{z}}_2^{k + 1})}
\end{array}} \right..
\end{equation}
Let $\{ {{\bf{v}}^{{k_j}}}\}$ be a convergent subsequence of $\{ {{\bf{v}}^k}\}$, since $\mathop {\lim }\limits_{k \to \infty } \| {{{\bf{v}}^{k + 1}} - {{\bf{v}}^k}} \|_2^2 = 0$, ${{\bf{v}}^{{k_j}}}$ and ${{\bf{v}}^{{k_j} + 1}}$ have the same limit point ${{\bf{v}}^*}: = ({\bf{x}}_{\rm{1}}^*,{\bf{x}}_2^*,{\bf{z}}_{\rm{1}}^*,{\bf{z}}_2^*,{\bf{w}}_{\rm{1}}^*,{\bf{w}}_2^*)$. Furthermore, since $\tilde L({\tilde{\bf{ v}}^k})$ is convergent, $\| {{\bf{z}}_1^{k + 1}} \|_{{q_1}}^{{q_1}}$ and $\| {{\bf{z}}_2^{k + 1}} \|_{{q_2}}^{{q_2}}$ are also convergent. Then, passing to the limit in (60) along the subsequence $\{ {{\bf{v}}^{{k_j}}}\}$ yields
\begin{equation} %
\begin{split}
&~~~~~~~~~~{\bf{x}}_1^* = {\bf{z}}_1^*,~~{\bf{x}}_2^* = {\bf{z}}_2^*,\\
&{\bf{w}}_1^* \in \beta \mu \partial \left\| {{\bf{z}}_1^*} \right\|_{{q_1}}^{{q_1}},~~{\bf{w}}_2^* \in \beta \partial \left\| {{\bf{z}}_2^*} \right\|_{{q_2}}^{{q_2}},\\
& - {\bf{w}}_1^* = 2{\bf{A}}_1^T({{\bf{A}}_1}{\bf{x}}_1^* + {{\bf{A}}_2}{\bf{x}}_2^* - {\bf{y}}),\\
& - {\bf{w}}_2^* = 2{\bf{A}}_2^T({{\bf{A}}_1}{\bf{x}}_1^* + {{\bf{A}}_2}{\bf{x}}_2^* - {\bf{y}}).\notag
\end{split}
\end{equation}
In particular, ${{\bf{v}}^*}$ is a stationary point of $\mathcal{L}$.

\textbf{Proof of Lemma 3:} Define ${\tilde{\bf{ v}}^k}: = ({\bf{x}}_{\rm{1}}^k,{\bf{x}}_2^k,{\bf{z}}_{\rm{1}}^k,{\bf{z}}_2^k,{\bf{w}}_{\rm{1}}^k,{\bf{w}}_2^k,{\bf{x}}_2^{k - 1})$, it follows from the definition of $\tilde {\mathcal{L}}$ that
\begin{equation}
{\partial _{{{\bf{z}}_1}}}\tilde {\mathcal{L}}({\tilde{\bf{ v}}^{k{\rm{ + 1}}}}) = \beta \mu \partial \left\| {{\bf{z}}_1^{k + 1}} \right\|_{{q_1}}^{{q_1}} - {\bf{w}}_1^{k + 1} - ({\bf{w}}_1^{k + 1} - {\bf{w}}_1^k)\notag
\end{equation}
which together with the first relation in (60) yields
\begin{equation}
{\rho _1}({\bf{x}}_1^k - {\bf{x}}_1^{k + 1}) + ({\bf{w}}_1^k - {\bf{w}}_1^{k + 1}) \in {\partial _{{{\bf{z}}_1}}}\tilde {\mathcal{L}}({\tilde{\bf{ v}}^{k{\rm{ + 1}}}}).\notag
\end{equation}
Similarly, we have
\begin{equation}
{\rho _2}({\bf{x}}_2^k - {\bf{x}}_2^{k + 1}) + ({\bf{w}}_2^k - {\bf{w}}_2^{k + 1}) \in {\partial _{{{\bf{z}}_2}}}\tilde {\mathcal{L}}({\tilde{\bf{ v}}^{k{\rm{ + 1}}}})\notag
\end{equation}
\begin{equation}
{\partial _{{{\bf{x}}_1}}}\tilde {\mathcal{L}}({\tilde{\bf{ v}}^{k{\rm{ + 1}}}}) = {\bf{w}}_1^{k + 1} - {\bf{w}}_1^k\notag
\end{equation}
\begin{equation}
{\partial _{{{\bf{x}}_2}}}\tilde {\mathcal{L}}({\tilde{\bf{ v}}^{k{\rm{ + 1}}}}) = ({\bf{w}}_2^{k + 1} - {\bf{w}}_2^k) + 2{c_1}{\rm{(}}{\bf{x}}_2^{k + 1} - {\bf{x}}_2^k{\rm{)}}\notag
\end{equation}
\begin{equation}
{\partial _{{\bf{\tilde x}}}}\tilde {\mathcal{L}}({\tilde{\bf{ v}}^{k{\rm{ + 1}}}}) = 2{c_1}{\rm{(}}{\bf{x}}_2^k - {\bf{x}}_2^{k + 1}{\rm{)}}\notag
\end{equation}
\begin{equation}
{\partial _{{{\bf{w}}_1}}}\tilde {\mathcal{L}}({\tilde{\bf{ v}}^{k{\rm{ + 1}}}}) = ({\bf{w}}_1^{k + 1} - {\bf{w}}_1^k)/{\rho _1}\notag
\end{equation}
\begin{equation}
{\partial _{{{\bf{w}}_2}}}\tilde {\mathcal{L}}({\tilde{\bf{ v}}^{k{\rm{ + 1}}}}) = ({\bf{w}}_2^{k + 1} - {\bf{w}}_2^k)/{\rho _2}.\notag
\end{equation}
Thus, there exists a constant ${c_5} > 0$ such that
\begin{equation}
\begin{split}
{\rm{dist}}(0,\partial \tilde {\mathcal{L}}({{\tilde{\bf{ v}}}^{k{\rm{ + }}1}})) &\le {c_5}\big({\left\| {{\bf{x}}_1^{k + 1} - {\bf{x}}_1^k} \right\|_2} + {\left\| {{\bf{x}}_2^{k + 1} - {\bf{x}}_2^k} \right\|_2}\\
 &~~~+ {\left\| {{\bf{w}}_1^{k + 1} - {\bf{w}}_1^k} \right\|_2} + {\left\| {{\bf{w}}_2^{k + 1} - {\bf{w}}_2^k} \right\|_2}\big)\notag
\end{split}
\end{equation}
which together with (42) and (44) results in Lemma 3. This result establishes a subgradient lower bound for the iterate gap, which together with Lemma 2 implies that
\begin{equation}
{\rm{dist}}(0,\partial \tilde {\mathcal{L}}({\tilde{\bf{ v}}^{k{\rm{ + }}1}})) \to 0 ~~{\rm{as}}~~k \to \infty.\notag
\end{equation}

\textbf{Proof of Theorem 2:} Based on the Lemma 2 and Lemma 3, the rest proof of Theorem 2 is to show the sequence $\{ {{\bf{v}}^k}\} $ generated via (18)--(23) has finite length i.e.,
\begin{equation}
\sum\limits_{k = 0}^\infty  {{{\left\| {{{\bf{v}}^{k + 1}} - {{\bf{v}}^k}} \right\|}_{\rm{2}}}}  < \infty
\end{equation}						
which means the sequence $\{ {{\bf{v}}^k}\}$ is a Cauchy sequence and thus is convergent. Consequently, this property together with Lemma 2 results in that the sequence $\{ {{\bf{v}}^k}\}$ globally converges to a critical point of $\mathcal{L}$. The property (61) can be derived based on the Kurdyka-Lojasiewicz (KL) property of $\tilde {\mathcal{L}}$. $\tilde {\mathcal{L}}$ is a KL function for arbitrary ${q_1} \ge 0$ and ${q_2} \ge 0$, since $\|\cdot\|_{{q_1}}^{{q_1}}$ and $\|\cdot\|_{{q_2}}^{{q_2}}$ are sub-analytic functions (thus KL functions) in this case. Since the detailed proof of (61) is similar to that for the 2-block ADMM in [40] with some minor changes, it is omitted here for succinctness.

\section{Proof of Theorem 3}
Define
\begin{equation}
f({\bf{x}}) = \left\| {\bf{x}} \right\|_{2}^q + \frac{\eta}{2}\left\| {{\bf{x}} - {\bf{t}}} \right\|_2^2.\notag
\end{equation}
For $q=0$, since $\left\| {\bf{x}} \right\|_2^0 = 0$ only when ${\bf{x}} = {\bf{0}}$ and $\left\| {\bf{x}} \right\|_2^0 = 1$ when ${\bf{x}} \ne {\bf{0}}$, it is easy to see that the minimizer of $f({\bf{x}})$, denoted by ${{\bf{x}}^ * }$, is given by
\begin{equation}
{{\bf{x}}^ * } = \left\{
\begin{aligned}
&{\bf{0}}, ~~~~~~{{{{\left\| {\bf{t}} \right\|}_2} < \sqrt {2/\eta } }}\\
&\{ {\bf{0}},{\bf{t}}\}, ~{{{\left\| {\bf{t}} \right\|}_2} = \sqrt {2/\eta } }\\
&{\bf{t}},~~~~~~~{\rm{otherwise}}
\end{aligned}
\right..
\end{equation}

For $0<q\leq 1$, by simple geometrical arguments, we first show that the minimizer ${{\bf{x}}^ * }$ satisfies that ${{\bf{x}}^ * } = \alpha {\bf{t}}$ with some $\alpha  \ge 0$. Specifically, assume that ${\left\| {{{\bf{x}}^ * } - {\bf{t}}} \right\|_2} = r$ and consider the set $\Omega  = \{ {\bf{x}}:{\left\| {{\bf{x}} - {\bf{t}}} \right\|_2} = r\}$, the points in the set $\Omega$ are lying on the ball with center at ${\bf{t}}$ and radius $r$. In the set $\Omega$, the minimal $\left\| \cdot \right\|_2^q$ value is given by the point which is the intersection of the ball and the vector ${\bf{t}}$. Thus, ${{\bf{x}}^ * } = \alpha {\bf{t}}$ with some $\alpha \ge 0$, with which we have
\begin{equation}
f({{\bf{x}}^ * }) = \left\| {\bf{t}} \right\|_2^q{\alpha ^q} + \frac{\eta }{2}\left\| {\bf{t}} \right\|_2^2{(\alpha  - 1)^2}.\notag
\end{equation}
Further, $\alpha$ should be the minimizer of the function $h(\alpha ) = \left\| {\bf{t}} \right\|_2^q{\alpha ^q} + \frac{\eta }{2}\left\| {\bf{t}} \right\|_2^2{(\alpha  - 1)^2}$. It has be shown in [37] that the minimizer of $h(\alpha )$ is given by $\alpha  = {\rm{pro}}{{\rm{x}}_{q,\eta \|{\bf{t}}\|{_2^{2 - q}}}}(1)$, which together with (62) results in (30) (can be computed via (7), (8) and (9)).

\section{ADMM Algorithm Applied to Problem (2)}

For the formulation (2), the standard 2-block ADMM procedure (S-ADMM) applies as follows [36]
\begin{equation}
\begin{split}
&{\bf{x}}_1^{k + 1} \\
&= \arg \mathop {\min }\limits_{{{\bf{x}}_1}} \left( {\mu \left\| {{{\bf{x}}_1}} \right\|_{{q_1}}^{{q_1}} + \frac{\rho }{2}\left\| {{{\bf{A}}_1}{{\bf{x}}_1} + {{\bf{A}}_2}{\bf{x}}_2^k - {\bf{y}} - \frac{{{{\bf{w}}^k}}}{\rho }} \right\|_2^2} \right)\notag
\end{split}
\end{equation}

\begin{equation}
\begin{split}
&{\bf{x}}_2^{k + 1} \\
&= \arg \mathop {\min }\limits_{{{\bf{x}}_2}} \left( {\left\| {{{\bf{x}}_2}} \right\|_{{q_2}}^{{q_2}} + \frac{\rho }{2}\left\| {{{\bf{A}}_1}{\bf{x}}_1^{k + 1} + {{\bf{A}}_2}{{\bf{x}}_2} - {\bf{y}} - \frac{{{{\bf{w}}^k}}}{\rho }} \right\|_2^2} \right)\notag
\end{split}
\end{equation}
\begin{equation}
{\bf{w}}_{}^{k + 1} = {\bf{w}}_{}^k - \rho ({{\bf{A}}_1}{\bf{x}}_1^{k + 1} + {{\bf{A}}_2}{\bf{x}}_2^{k + 1} - {\bf{y}})\notag
\end{equation}
Both the ${{\bf{x}}_1}$- and ${{\bf{x}}_2}$-subproblems are ${\ell _q}$-regularized least-square problem which are difficult to solve directly. The standard trick is to adopt a proximal linearization of each subproblem. Let ${{\bf{u}}^k} = {{\bf{A}}_2}{\bf{x}}_2^k - {\bf{y}} - {{\bf{w}}^k}/\rho $, consider a quadratic majorization of the second term in the ${{\bf{x}}_1}$-subproblem as
\begin{equation}
\begin{split}
&\frac{1}{2}\left\| {{{\bf{A}}_1}{{\bf{x}}_1} + {{\bf{u}}^k}} \right\|_2^2\\
& \approx \frac{1}{2}\left\| {{{\bf{A}}_1}{\bf{x}}_1^k + {{\bf{u}}^k}} \right\|_2^2 + \left\langle {{{\bf{x}}_1} - {\bf{x}}_1^k,{g_1}({\bf{x}}_1^k)} \right\rangle  + \frac{{{c_1}}}{2}\left\| {{{\bf{x}}_1} - {\bf{x}}_1^k} \right\|_2^2\notag
\end{split}
\end{equation}
where ${g_1}({\bf{x}}_1^k) = {\bf{A}}_1^T({{\bf{A}}_1}{\bf{x}}_1^k + {{\bf{u}}^k})$, ${c_1} > 0$ is a proximal parameter. Then, the ${{\bf{x}}_1}$-subproblem becomes a form of the ${\ell _q}$-norm proximity operator as
\begin{equation}
{\bf{x}}_1^{k + 1} = {\rm{pro}}{{\rm{x}}_{{q_1},{c_1}\rho /\mu }}\left( {{\bf{x}}_1^k - {g_1}({\bf{x}}_1^k)/{c_1}} \right). \notag
\end{equation}
Similarly, the ${{\bf{x}}_2}$-subproblem can be solved as
\begin{equation}
{\bf{x}}_2^{k + 1} = {\rm{pro}}{{\rm{x}}_{{q_2},{c_2}\rho }}\left( {{\bf{x}}_2^k - {g_2}({\bf{x}}_2^k)/{c_2}} \right)\notag
\end{equation}
where ${g_2}({\bf{x}}_2^k) = {\bf{A}}_2^T({{\bf{A}}_1}{\bf{x}}_1^{k + 1} + {{\bf{A}}_2}{\bf{x}}_2^k - {\bf{y}} - {{\bf{w}}^k}/\rho )$ and ${c_2} > 0$.

\end{document}